\documentclass[12pt,preprint]{aastex}

\slugcomment{Accepted for publication in the ApJ}

\shorttitle{The Clustering of EROs}
\shortauthors{Brown {\it et al.}}

\begin{document}

\title{The Clustering of Extragalactic Extremely Red Objects}

\author{Michael J. I. Brown\altaffilmark{1,2}, Buell T. Jannuzi\altaffilmark{2}, Arjun Dey\altaffilmark{2}, Glenn P. Tiede\altaffilmark{3}}
\altaffiltext{1}{Department of Astrophysical Sciences, Princeton University, Peyton Hall, Princeton, NJ 08544-1001.}
\altaffiltext{2}{National Optical Astronomy Observatory, Tucson, AZ 85726-6732}
\altaffiltext{3}{Department of Physics and Astronomy, 104 Overman Hall, Bowling Green State University, Bowling Green, OH 43403}
\email{mbrown@astro.princeton.edu, jannuzi@noao.edu, dey@noao.edu, gptiede@bgnet.bgsu.edu}

\begin{abstract}
We have measured the angular and spatial clustering of 671 $K<18.40$, $R-K>5$ Extremely 
Red Objects (EROs) from a $0.98~{\rm deg}^2$ sub-region of the NOAO Deep Wide-Field Survey (NDWFS). 
Our study covers nearly 5 times the area and has twice the sample size of
any previous ERO clustering study.
The wide field of view and $B_WRIK$ passbands of the NDWFS allow us to place improved 
constraints on the clustering of $z\sim 1$ EROs. We find the angular clustering
of EROs is slightly weaker than in previous measurements, and 
$\omega(1^\prime)=0.25 \pm 0.05$ for $K<18.40$ EROs. 
We find no significant correlation of ERO spatial clustering with redshift, 
apparent color or absolute magnitude, although given the uncertainties, 
such correlations remain plausible. We find the spatial clustering of 
$K<18.40$, $R-K>5$ EROs is well approximated by a power-law, with 
$r_0=9.7 \pm 1.1 ~h^{-1} {\rm Mpc}$ in comoving coordinates.
This is comparable to the clustering of $\sim 4L^*$ early-type galaxies at $z<1$,
and is consistent with the brightest EROs being the progenitors of the most massive ellipticals.
There is evidence of the angular clustering of EROs decreasing with increasing
apparent magnitude, when NDWFS measurements of ERO clustering are combined with 
those from the literature. Unless the redshift distribution of 
$K \gtrsim 20$ EROs is very broad, the spatial clustering of EROs decreases from 
$r_0=9.7 \pm 1.1 ~h^{-1} {\rm Mpc}$ for $K<18.40$ to $r_0\sim 7.5 ~h^{-1}{\rm Mpc}$
for $K \gtrsim 20$ EROs.
\end{abstract}


\section{Introduction}
\label{sec:intro}

The evolution of galaxy clustering is a prediction of hierarchical
models of galaxy and structure formation \citep[e.g.,][]{kau99,ben01,som01}. 
Hierarchical models for a concordance cosmology\footnote{Throughout this paper 
$H_0 \equiv 100~h~ {\rm km~s}^{-1} ~{\rm Mpc}^{-1}$, $\Omega_m=0.3$, 
$\Lambda=0.7$, and comoving coordinates.} predict little or no evolution of the clustering 
of $\gtrsim L^*$ red galaxies at $z<2$. Precise measurements of 
galaxy clustering at $z\sim 1$ can therefore test the predictions of these models.

Extremely Red Objects \citep[EROs;][]{els88,mcc92,hu94,dey95} could be the 
progenitors of local ellipticals \citep[e.g.,][]{spi97}. Roughly 80\% of $K_S<18.7$ EROs 
have spectra with the absorption features of old stellar populations \citep{yan04}
and $\sim50\%$ of $K\lesssim 22$ EROs have early-type morphologies \citep{mor00,sti01,mou04}.
Some EROs contain super-massive black holes, as $\sim 15\%$ of EROs contain
an obscured Active Galactic Nucleus which can be detected by deep X-ray 
surveys \citep{ale02,roc03}. A direct test of the relationship between $z\sim 1$ EROs and
the most massive local ellipticals is to compare the spatial clustering of the
two populations.

Previous constraints on the spatial correlation function of EROs,
summarized in Table~\ref{table:prev}, are provided by 
pencil-beam surveys with $\lesssim 0.2~{\rm deg}^2$ areal coverage each.
Individual structures comprised of EROs can have sizes comparable to the 
field of view of these surveys \citep[e.g.,][]{dad00}, 
and small surveys do not sample representative
volumes of the Universe for highly clustered objects \citep[e.g.,][]{som04}. 
At $z\sim 1$, the transverse comoving distance spanned by 
previous ERO studies is $\lesssim 20 ~h^{-1} {\rm Mpc}$, which is much smaller than 
the size of individual structures observed in the present-day Universe.
Spatial clustering measurements derived 
from the angular correlation function depend on ERO redshift 
distribution models. Previous angular clustering studies were unable 
to verify their model redshift distributions, as complete 
spectroscopic samples of EROs were unavailable. Previous ERO spatial 
clustering measurements have large uncertainties and possibly large 
(and sometimes unaccounted for) systematic errors.

In this paper, we present a measurement of the clustering of EROs using
$B_WRIK$ imaging of a $0.98~{\rm deg}^2$ subset of the NOAO 
Deep Wide-Field Survey (NDWFS). The large area of our study
provides a more representative volume than previous studies. 
The $B_WRIK$ passbands of the NDWFS
allow us to constrain the ERO redshift distribution with photometric 
redshifts and their uncertainties. We also use photometric redshifts to 
select EROs as a function of luminosity and redshift. We use 
ERO spectroscopic redshifts to verify the accuracy of our photometric 
redshifts and we compare our estimate of the ERO redshift distribution 
with spectroscopic redshift distributions from the literature.

The structure of the paper is as follows.
In \S\ref{sec:ndwfs} we provide a brief description of the NDWFS imaging 
and catalogs from which the $K<18.40$ ERO
sample was selected. We discuss our estimates of ERO photometric redshifts, 
and provide a comparison of ERO photometric and spectroscopic
redshifts in \S\ref{sec:photoz}. The selection of the ERO sample and ERO number
counts are discussed in \S\ref{sec:ero}.  In \S\ref{sec:cor}, we describe 
the techniques used to measure the angular and spatial correlation functions.
The angular and spatial clustering of EROs, as a function of apparent magnitude, apparent color,  
absolute magnitude, and redshift are discussed in \S\ref{sec:clu}. We discuss the 
implications of our results in \S\ref{sec:dis} and summarize the paper in 
\S\ref{sec:sum}.

\section{The NOAO Deep Wide-Field Survey}
\label{sec:ndwfs}

The NDWFS is a multiband ($B_W,R,I,K$) survey of two
$\approx 9.3~{\rm deg}^2$ high Galactic latitude fields with the CTIO
$4{\rm m}$, KPNO $4 {\rm m}$, and KPNO $2.1 {\rm m}$ telescopes \citep{jan99}.
A thorough description of the optical and $K$-band observing strategy and data reduction
will be provided by Jannuzi et al. and Dey et al. (both in preparation). 
This paper utilizes $0.98~{\rm deg}^2$ of $B_WRIK$ data in the Bo\"{o}tes field.
$B_WRI$ imaging and catalogs for the entire NDWFS  Bo\"{o}tes field became available
from the NOAO Science Archive\footnote{http://www.archive.noao.edu/ndwfs/} on 22 October 2004.
$K$-band imaging and catalogs for approximately half of the Bo\"{o}tes field are 
also available from this archive.

We generated object catalogs using SExtractor $2.3.2$ \citep{ber96},
run in single-image mode in a similar manner to \cite{bro03}.
At faint magnitudes, detections in the different bands were matched
if the centroids were within  $1^{\prime\prime}$ of each other.
At bright magnitudes, detections in the different bands were matched
if the centroids were within an ellipse defined using the second
order moments of the light distribution of the object\footnote{This 
ellipse was defined with the SExtractor parameters $2 \times 
{\rm A\_WORLD}$, $2 \times {\rm B\_WORLD}$, and ${\rm THETA\_WORLD}$.}.
Throughout this paper we use SExtractor MAG\_AUTO magnitudes
\citep[which are similar to Kron total magnitudes;][]{kro80}, due to their
small uncertainties and systematic errors at faint magnitudes.
Our clustering measurements are not particularly sensitive to how
we measure ERO photometry, and the clustering of EROs selected
with $4^{\prime\prime}$ diameter aperture photometry is only marginally
weaker than the clustering of EROs selected with MAG\_AUTO photometry.

We determined the completeness as a function of magnitude by adding artificial
objects to copies of the data and recovering them with SExtractor.
To approximate $z\sim 1$ galaxies, the artificial objects have an 
intrinsic profile with a full width at half maximum of $0.5^{\prime\prime}$,
which was then convolved with a Moffat profile model of the seeing.
The $50\%$ completeness limits vary within the sample area in the ranges of
$26.0<B_W<26.7$, $24.8<R<25.6$, $23.6<I<25.2$ and 
$18.6<K<18.7$\footnote{Throughout this paper we use Vega photometry.}. 

Regions surrounding saturated stars were removed from the catalog to
exclude (clustered) spurious objects detected in the wings of the point
spread function. We excluded regions where the root-mean-square of the
sky noise in the $K$-band data was $20\%$ higher than the mean, 
as the depth of these regions is significantly less than the mean depth across 
the field. While it is plausible that smaller variations in the sky noise 
could alter the measured clustering of the faintest EROs, our main conclusions
remain unchanged if we exclude $K>18.15$ EROs from the sample.

We used SExtractor's star-galaxy classifier to remove objects from the 
galaxy catalog which had a stellarity of $>0.7$ in 2 or more bands 
brighter than $B_W<23.8$, $R<22.8$, and $I<21.4$. At fainter magnitudes
we do not use the star-galaxy classification and correct the angular correlation
function for the estimated stellar contamination of the sample. We do not use the 
$K$-band for star-galaxy classification as there are image quality variations 
across the $K$-band image stacks. 
We estimated stellar contamination of the galaxy sample 
using the same technique as \cite{bro03}, where the stellar number counts
were assumed to be a power-law and the distribution of stellar colors 
does not change with magnitude at $R\gtrsim 21$. The contamination of 
the ERO sample (\S\ref{sec:ero}) by stars is estimated to be $\sim 2\%$, and 
the conclusions of this paper remain unaltered unless stellar contamination 
is higher than $15\%$.

\section{Photometric Redshifts}
\label{sec:photoz}

\begin{figure}
\plottwo{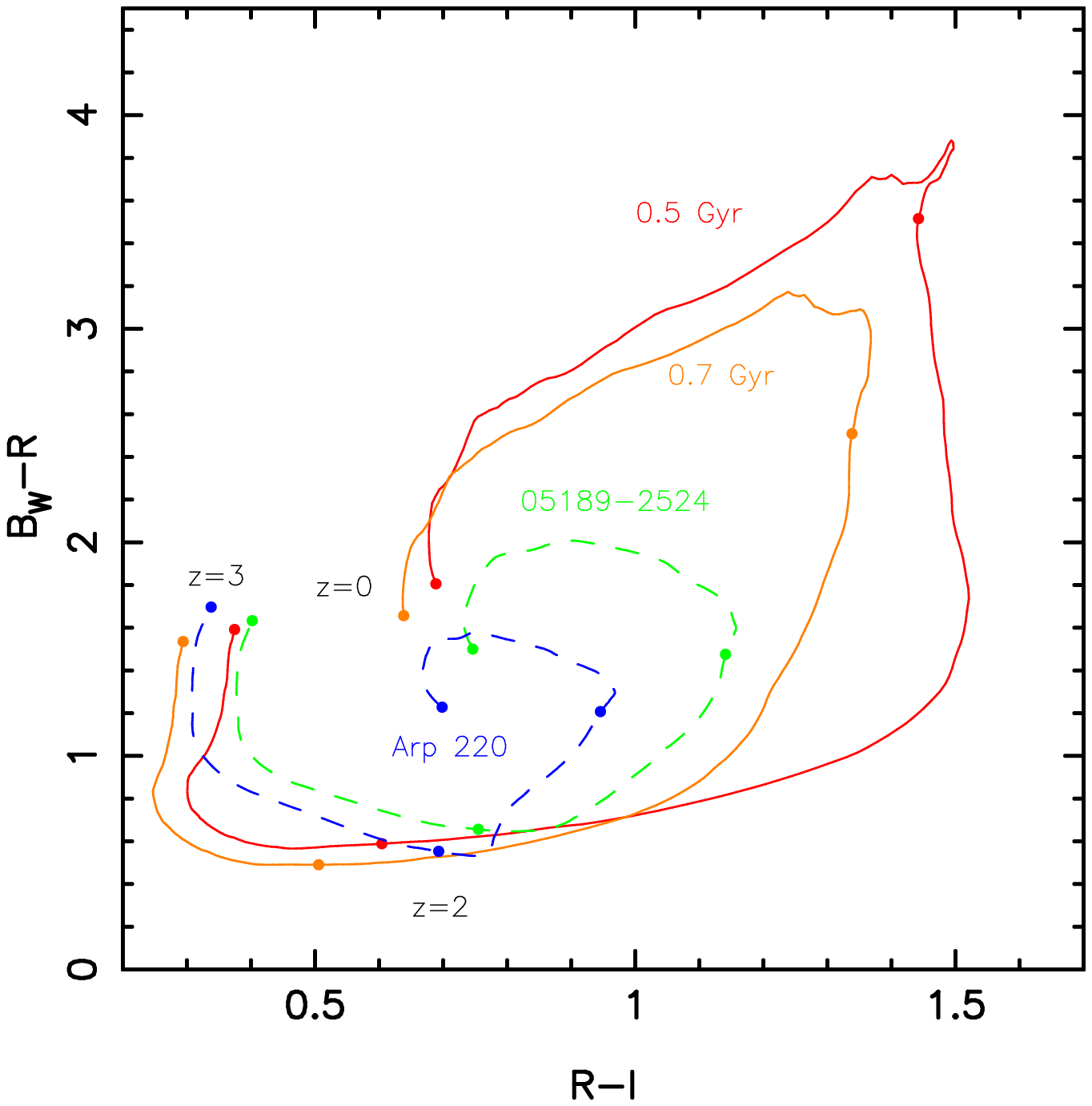}{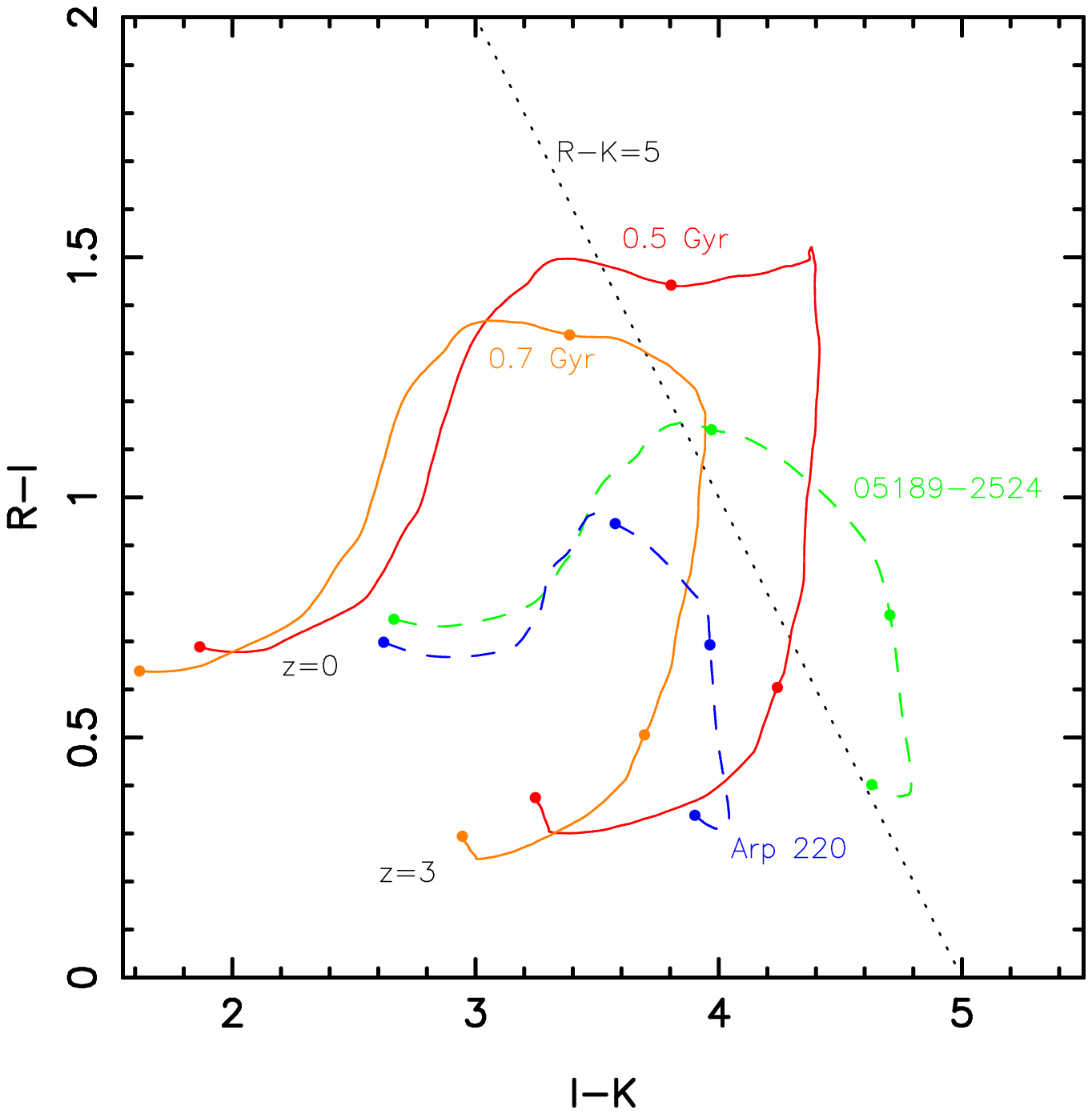}
\caption{Color-color diagrams of the PEGASE2 $\tau$-models and 
non-evolving templates of Arp 220 and IRAS 05189-2524 \citep{dev99}. 
The $R-K=5$ selection criterion for EROs is shown in the right-hand panel. 
Dots mark $z=0$, $1$, $2$, and $3$ on the model tracks. For clarity, the 
redshift range shown is restricted to $0\leq z\leq 3$ and we have not plotted 
$\tau$-models bluer than the ERO color cut. 
\label{fig:colmod}}
\end{figure}

Photometric redshifts were determined for all objects with $I$ 
and $K$-band detections. We provide a brief overview of the photometric
redshifts here and refer the reader to our earlier study of 
$0.3<z<0.9$ red galaxy clustering in the NDWFS \citep{bro03}
for a more detailed description of the photometric redshift code. 
To model galaxy spectral energy distributions (SEDs), we used 
PEGASE2 evolutionary synthesis models \citep{fio97} 
with exponentially declining star formation rates ($\tau$ models) and $z=0$ 
ages of $12 ~{\rm Gyr}$ (formation $z \approx 4$). 
The effect of $E(B-V)=0.04$ dust reddening with $R_V=3.1$, comparable to estimates for
$0<z<1$ early-type galaxies \citep{fal99}, was included in the $\tau$ models. 
In \cite{bro03},
we used models with solar metallicity at $z=0$, which resulted in small systematic underestimates of 
galaxy redshifts. Simple solar metallicity $\tau$ models 
underestimate the UV luminosity of galaxies \citep[e.g.,][]{don95}, so in this
work we let the metallicity of the models be a function of $\tau$.
This has the effect of slightly increasing the UV flux of the model
SEDs. We verified the accuracy of the photometric redshifts at $z<1$ with 
89 $M_R<-19-5{\rm log}h$ galaxies with rest-frame $B_W-R>1.05$ and spectroscopic redshifts. 
After decreasing the metallicity of the models, the photometric redshifts of these 
red galaxies did not have significant systematic errors.
We note, however, that the UV flux in galaxies can also be increased 
by the presence of young stars or by altering the properties of the dust extinction; 
our approach is merely a proxy for correcting any systematic 
effects in our photometric redshifts and is not meant to be 
interpreted as justifying sub-solar metallicities in the red galaxy population.
We use these solar and sub-solar $\tau$ models throughout the remainder of the paper.
Color-tracks for 2 of the models are shown in Figure~\ref{fig:colmod}.
For comparison, we also show two ultra-luminous infrared galaxy (ULIRG) 
templates from \cite{dev99}, which have bluer $B_W-R$ colors than the 
$\tau$ models at $z\sim 1$.

Photometric redshifts were estimated by finding the minimum value of $\chi^2$
as a function of redshift, spectral type ($\tau$), and luminosity.
For objects not detected in the $R$ or $B_W$-bands, we estimated the probability 
of a non-detection using the completeness estimates 
discussed in \S\ref{sec:ndwfs}. As the model SEDs do not account for the 
observed width of the galaxy locus, we increased the photometric 
uncertainties for the galaxies by $0.05$ magnitudes (added in quadrature).
To improve the accuracy of the photometric redshifts, 
the estimated redshift distribution of galaxies as a function of
spectral type and apparent magnitude was introduced as a prior.
The 2dF Galaxy Redshift Survey (2dFGRS) luminosity functions for different spectral types
\citep{mad02}, with spectral evolution given by the $\tau$-models, were used to
estimate the redshift distributions.

We tested the reliability of the photometric redshifts with simulated
galaxies and real galaxies with spectroscopic redshifts. Simulated galaxies
were generated using the PEGASE2 $\tau$ models.  The simulated data 
consisted of $K<18.40$ galaxies with $0.6~{\rm Gyr} \leq \tau \leq 15~{\rm Gyr}$ 
in the redshift range $0<z\leq5$ and luminosity range $0.01<L^*\leq 100$.
The simulated object photometry was scattered using the estimated
uncertainties, thus mimicking what would be present in the real catalogs.
We tested the accuracy of the photometric redshifts
with a few spectroscopic redshifts and $B_WRIK$ photometry for
EROs in the NDWFS Bo\"{o}tes field.

\begin{figure}
\plottwo{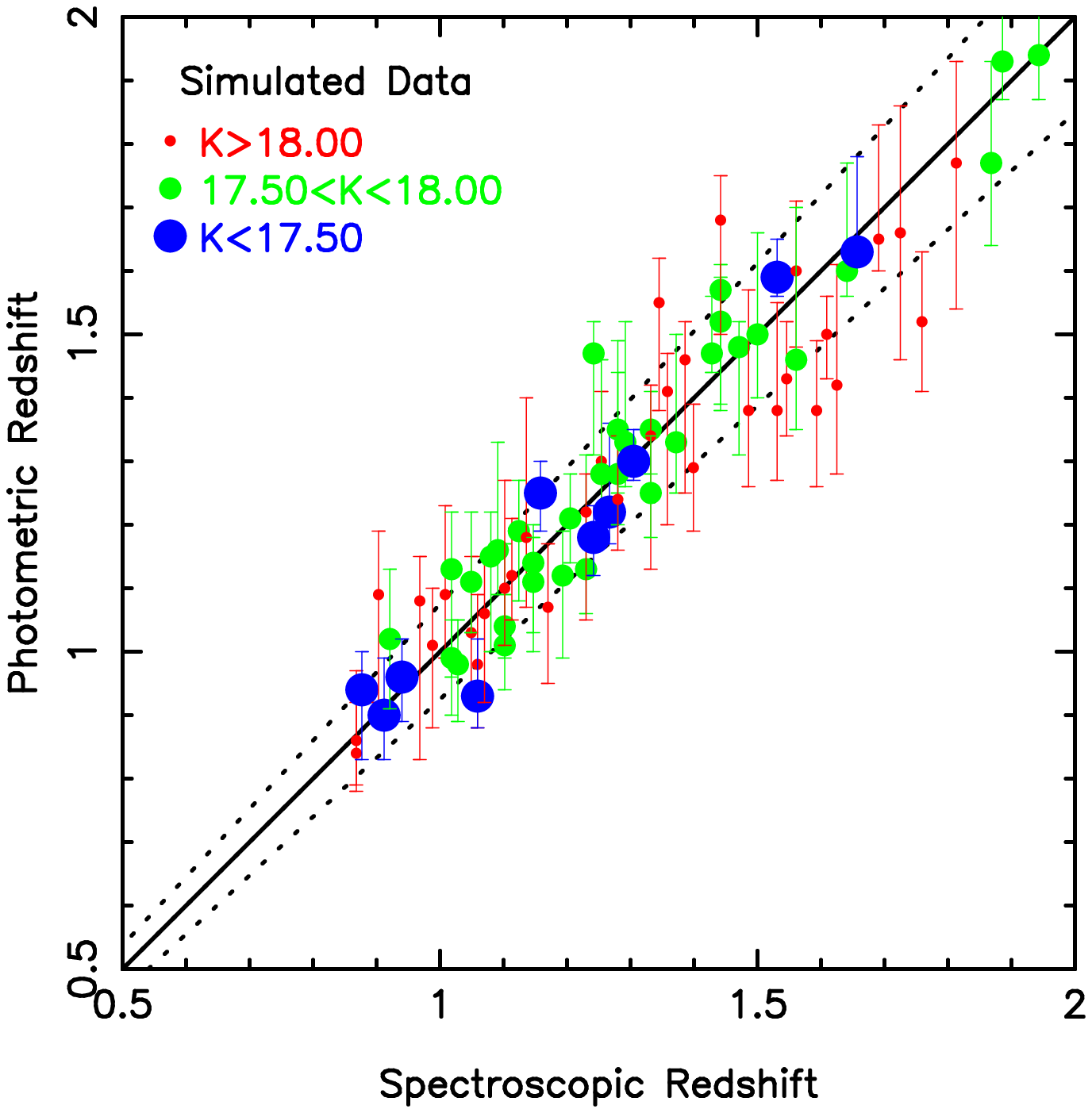}{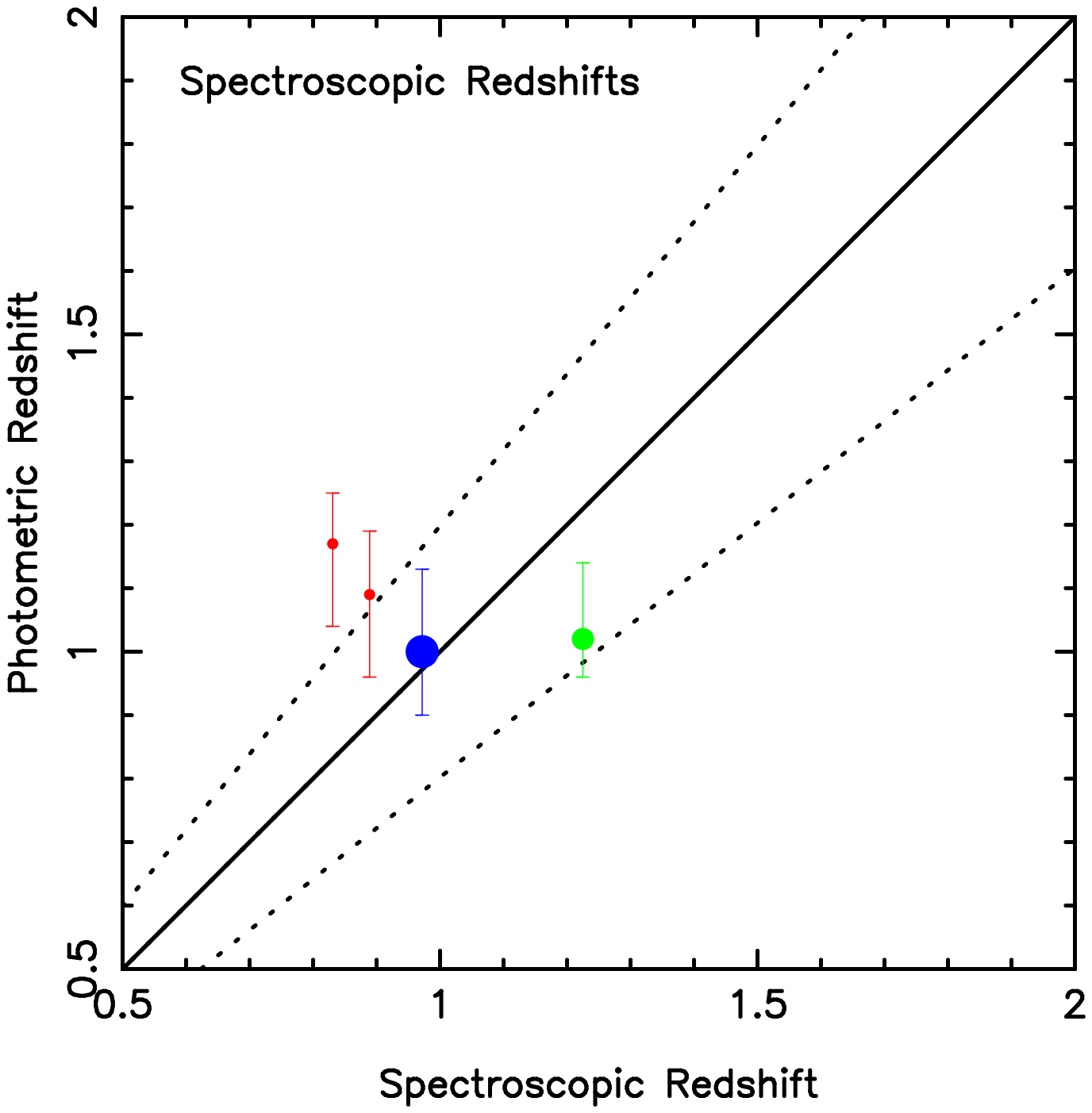}
\caption{A comparison of $K<18.40$ $R-K>5.0$ ERO photometric and spectroscopic 
redshifts for simulated and real data. Dotted diagonal lines show the measured 
$\pm 1 \sigma$ uncertainties of the photometric redshifts. Simulated EROs, generated 
using the $\tau$-models with photometric noise added, are
shown in the left-hand panel. On the right are real EROs with spectroscopic
redshifts. For clarity only a third of the simulated galaxies are plotted. The measured
$1\sigma$ uncertainties of the photometric redshifts for real EROs is $\sim 20\%$.
\label{fig:photoz}}
\end{figure}

A comparison of our photometric and spectroscopic redshifts for
EROs is shown in Figure~\ref{fig:photoz}. We discuss the selection 
criteria for the EROs in \S\ref{sec:ero}. The simulated galaxies 
in the left panel of Figure~\ref{fig:photoz} have $1\sigma$ uncertainties 
of $\simeq 8\%$. For galaxies with SEDs similar to the PEGASE2 $\tau$ models,
our procedure should yield accurate photometric redshifts.
There are 4 NDWFS $K<18.40$ EROs with spectroscopic redshifts. 
As shown in the right hand panel of Figure~\ref{fig:photoz}, real EROs in 
the NDWFS exhibit a $1\sigma$ scatter of $\sim 20\%$ between the photometric and 
spectroscopic redshifts. There are more outliers than would
be expected if the $\tau$ models reproduced the variety of ERO SEDs. 
For comparison, GOODS obtains ERO photometric redshifts with 
accuracies of $\sim 5\%$ \citep{mob04}, as they have photometry and upper
limits in more bands ($U^{\prime}UBB_{435}VV_{606}RIi_{775}z_{850}JHK_S$).
The accuracy of photometric redshifts is a complex function of redshift, SED and 
apparent magnitude. The accuracy of the ERO photometric redshifts could not be 
extrapolated from $z<1$ red galaxy photometric redshifts, which can have 
uncertainties of $<10\%$ \citep[e.g.,][]{bro03}. The accuracy of ERO photometric redshifts can 
not, and should not, be extrapolated from other samples of galaxies, such as 
samples selected by apparent magnitude only or from the Hubble Deep Fields (HDFs). 
Though our ERO photometric redshifts can only be considered approximations, they provide 
a good estimate of the ERO redshift distribution (see \S\ref{sec:cor}).

\section{The Extremely Red Object Sample}
\label{sec:ero}

\begin{figure}
\plottwo{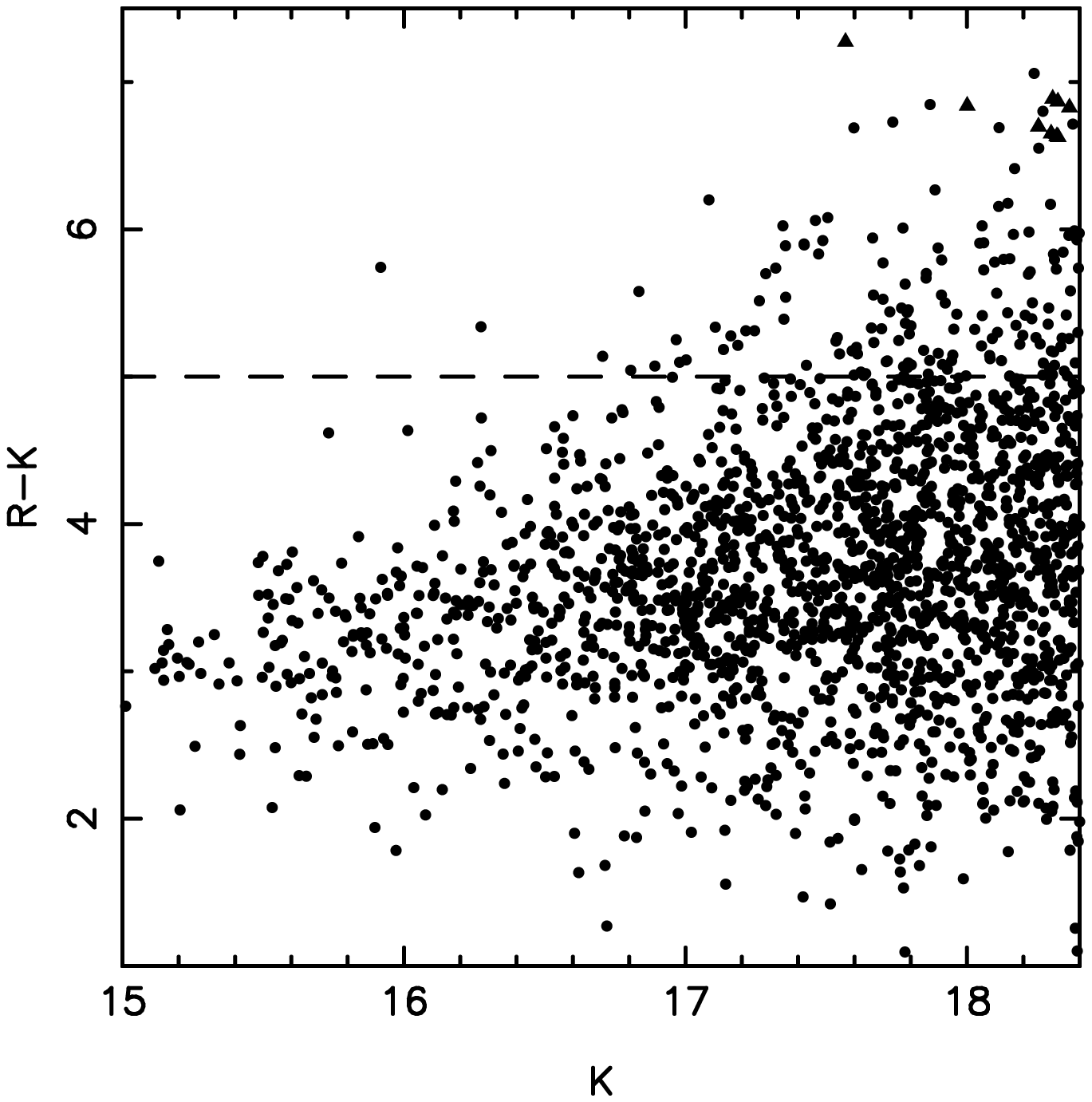}{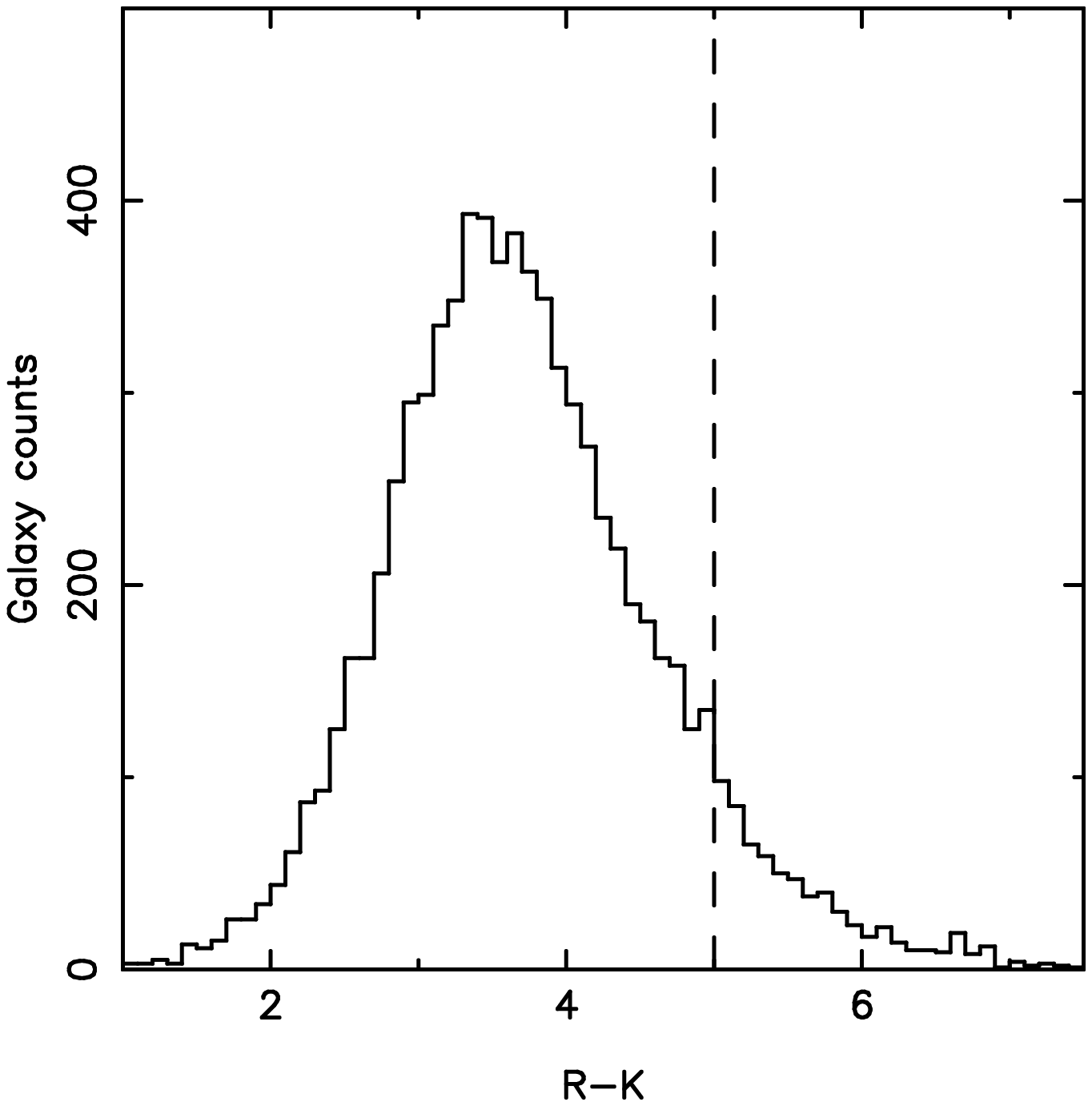}
\caption{The color-magnitude diagram and the number of $K<18.40$ galaxies
as a function of $R-K$ color. The $R-K=5$ selection criterion is marked with dashed lines.
Only 25\% of all the galaxies have been plotted in the color-magnitude diagram
and $R-K$ lower limits are denoted by triangles. The $R$ magnitude limits vary
across the $0.98~{\rm deg}^2$ sub-region, as the $R$-band imaging consists of 4 pointings.
The uncertainty of the galaxy colors increases from $\simeq 0.1$ 
magnitudes at $K=17.00$ to $\simeq 0.25$ magnitudes at $K=18.40$.  
$R-K>5.0$ EROs comprise $\simeq 8\%$ of all $K<18.4$ galaxies.
\label{fig:colmag}}
\end{figure}

We selected EROs with the $R-K>5$ 
criterion \citep[e.g.,][]{els88,dad00,roc02}, though redder color cuts are 
sometimes used in the literature \citep[e.g.,][]{hu94,dey99}. 
We have limited the sample to $K<18.40$ EROs, to reduce the effects
of completeness variations across the survey area on the measured clustering.
As shown in Figure~\ref{fig:colmag}, the percentage of EROs increases from 
$\simeq 0\%$ of the total galaxy counts at $K<16$ 
to $\simeq 8\%$ at $K\simeq 18.4$. 

Contamination of the ERO sample by other galaxies could significantly alter the 
measured correlation function. At the magnitude limit of our sample, the uncertainty 
in the $R-K$ color is $\simeq 0.25$ magnitudes. For the distribution of galaxy 
colors shown in Figure~\ref{fig:colmag}, and assuming Gaussian photometric uncertainties, 
approximately 6\% of the $K<18.40$ ERO sample is contamination by $R-K<4.75$ galaxies.
Even if $R-K<4.75$ galaxies were completely (and implausibly) unclustered, the amplitude of the 
$R-K>5$ angular correlation function would only be decreased by 12\%.
Contamination by $4.75<R-K<5.00$ galaxies could be as high as 22\% in the 
$K<18.40$ ERO sample. This would significantly alter our results if the clustering
of galaxies is a very strong function of color at $R-K\sim 5$. However, as discussed in 
\S\ref{sec:color}, we do not see evidence of this within our dataset.
\cite{mal20} bias does increase the observed number of EROs. If we assume the ERO number
counts in Table~\ref{table:ang} are a good approximation of the true ERO number counts,
then the contribution of Malmquist bias to the NDWFS counts is $\lesssim 8\%$.
This would alter the measured clustering if ERO angular clustering is an extremely 
strong function of apparent magnitude.

\begin{figure}
\plottwo{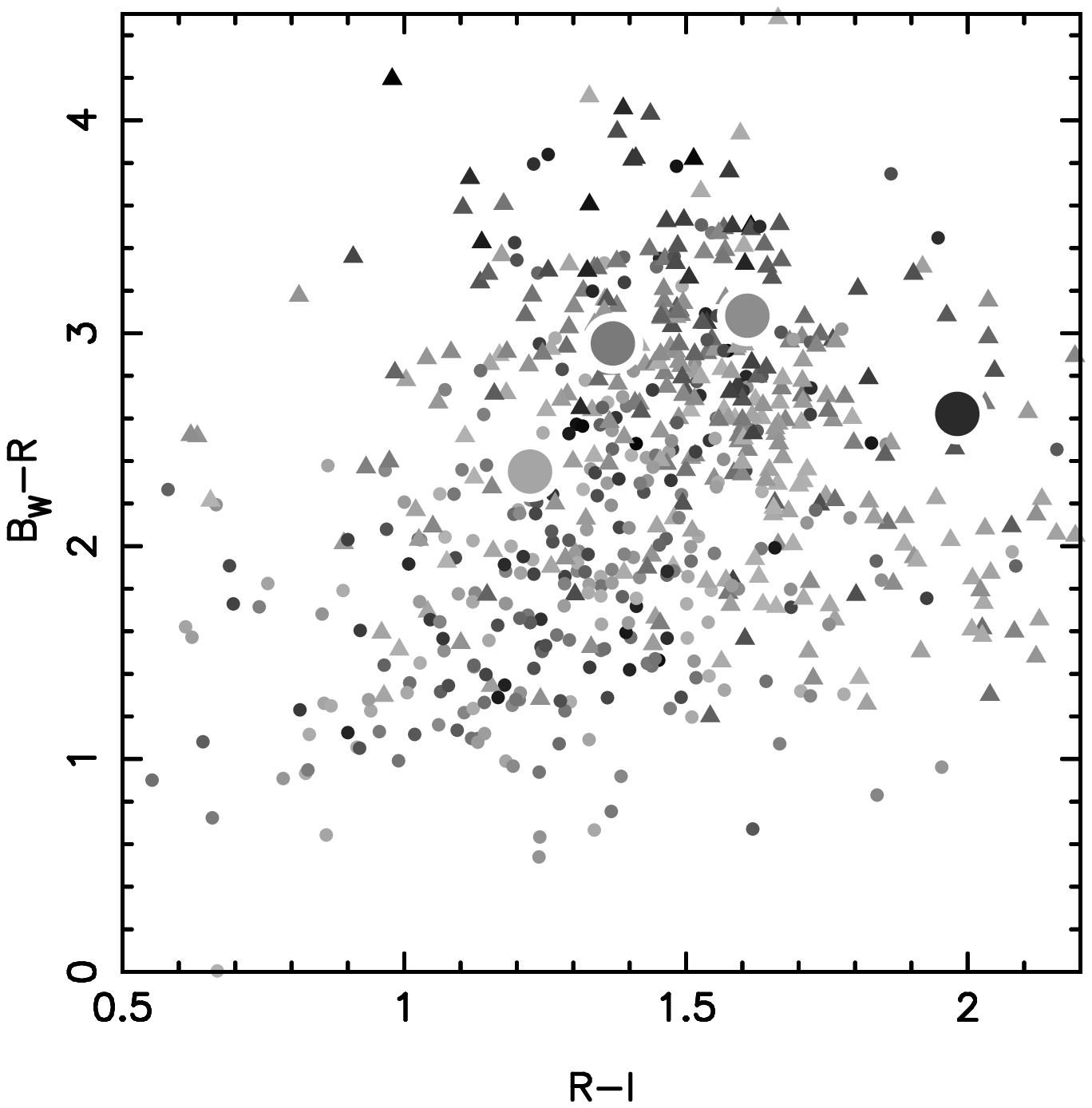}{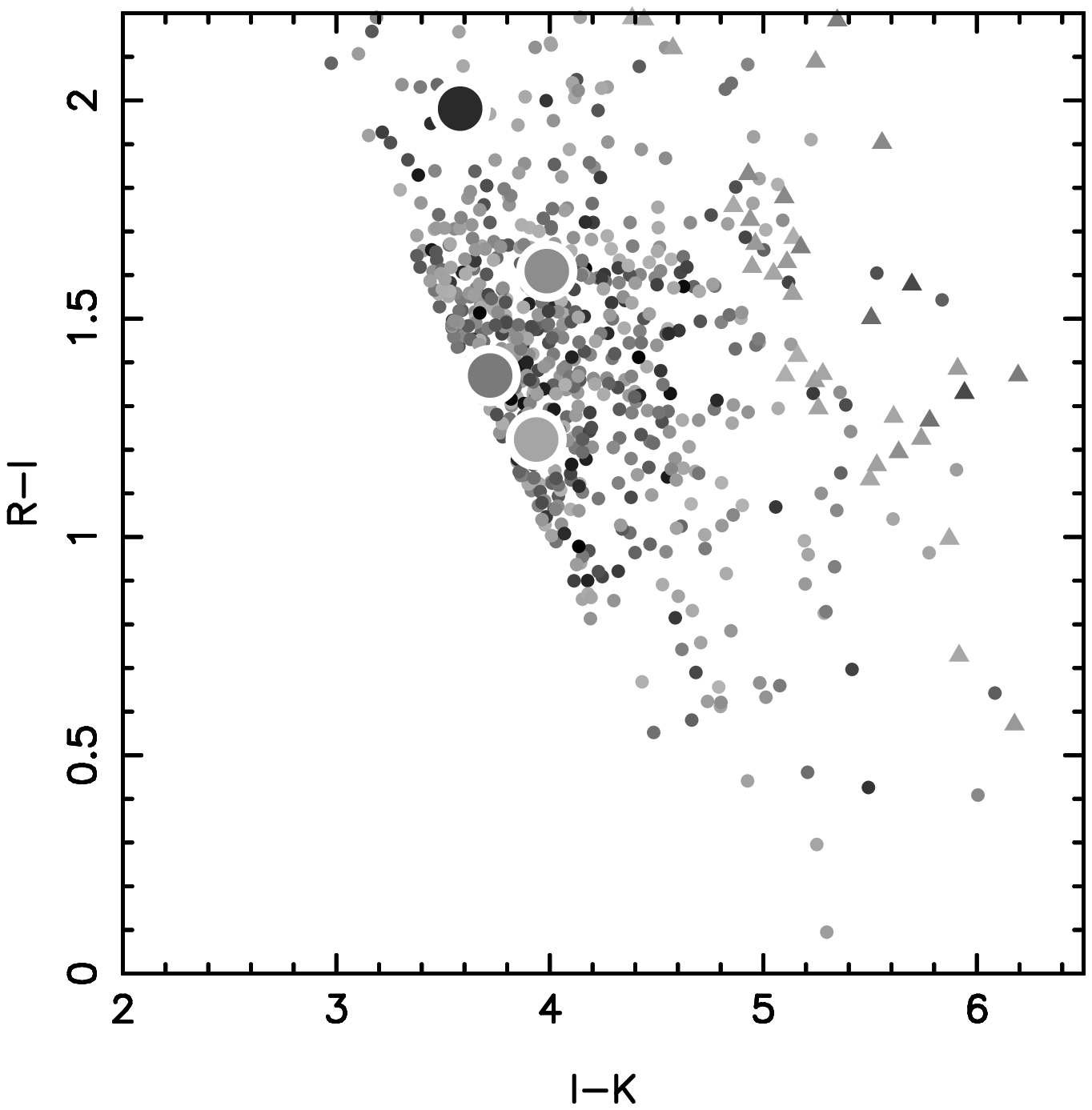}
\caption{Color-color diagrams of the NDWFS ERO sample. $B_W$ (left panel) and $R$ (right panel)
non-detections are shown with triangles and large symbols denote EROs with spectroscopic redshifts.
Black symbols are $K=17$ or brighter while paler symbols are fainter.
The $B_W-R$ colors of most $K<18.40$ EROs are redder than the ULIRG templates 
plotted in Figure~\ref{fig:colmod}. A broad locus of galaxies can be seen at $R-I\sim 1.5$,
which is coincident with the reddest PEGASE2 $\tau$ models at $1.0<z<1.6$. 
The faintest objects in the sample have photometric uncertainties of $\sim 0.25$ magnitudes,
so some objects with very unusual colors may be photometric errors. 
\label{fig:colcol}}
\end{figure}

We assume the bulk of our sample consists of galaxies with red stellar populations.
The colors of dusty starbursts are predicted to differ significantly from galaxies 
with red stellar populations. As shown in Figure~\ref{fig:colcol}, 77\% of the NDWFS 
ERO sample has $R-I>1.15$, which is redder than the \cite{dev99} non-evolving ULIRG templates shown in 
Figure~\ref{fig:colmod}. Our assumption that most $K<18.40$ EROs have red 
stellar populations is also consistent with the conclusions of \cite{yan04}, who find 
86\% of $K_S< 18.7$ EROs have the absorption features of old stellar populations.

\begin{figure}
\plotone{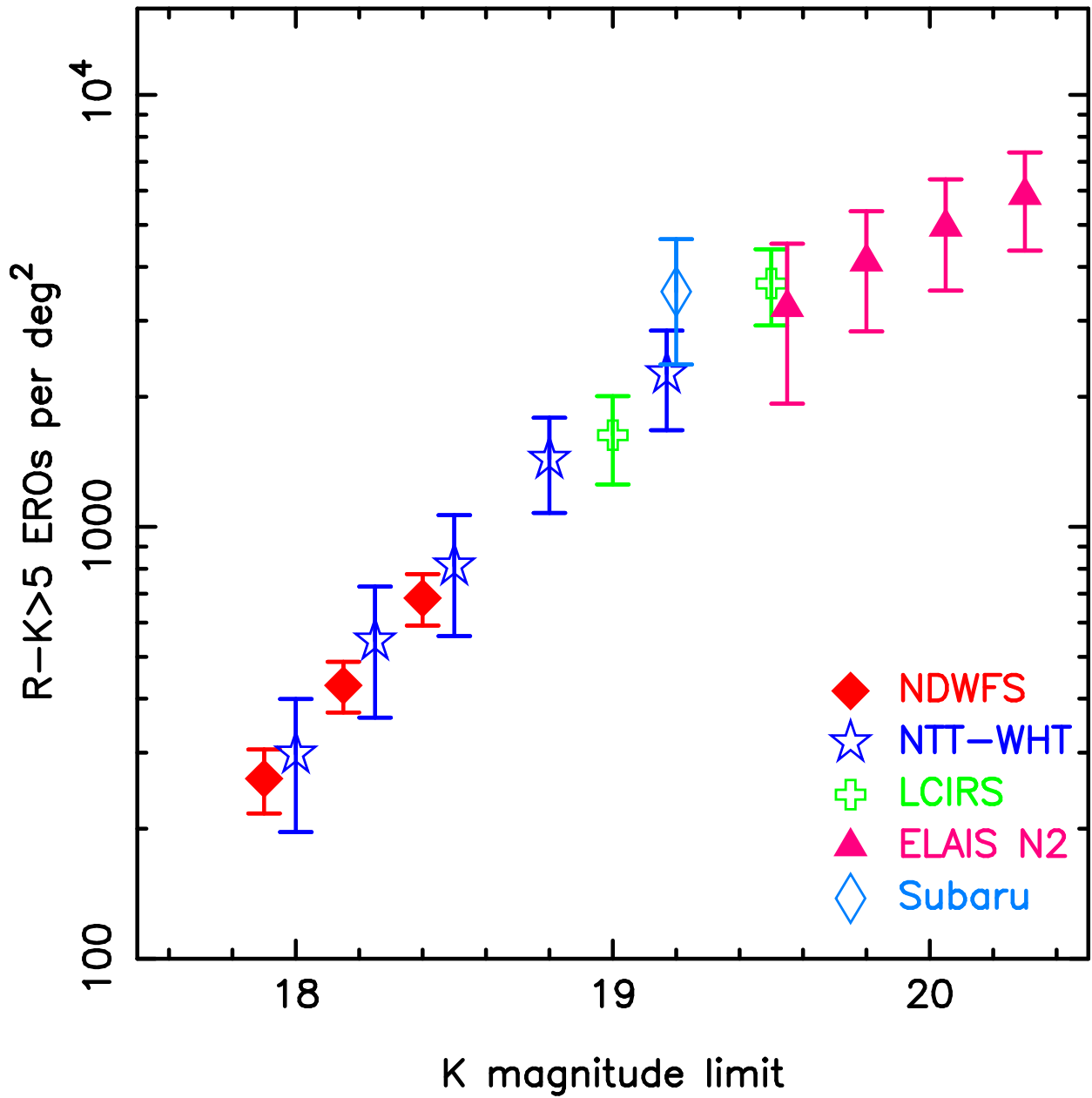}
\caption{The cumulative sky surface density of $R-K>5$ EROs as a function 
of $K$-band magnitude limit. LCIRS $R-H>4$ EROs have been included using the assumption $H-K=1$.
References for each survey are listed in Table~\ref{table:prev}.
We have estimated the uncertainties for each study using the integral constraint 
and assumption of square fields of view. The sky surface density of EROs measured 
by the different surveys are in good agreement.
\label{fig:counts}}
\end{figure}

\begin{figure}
\plotone{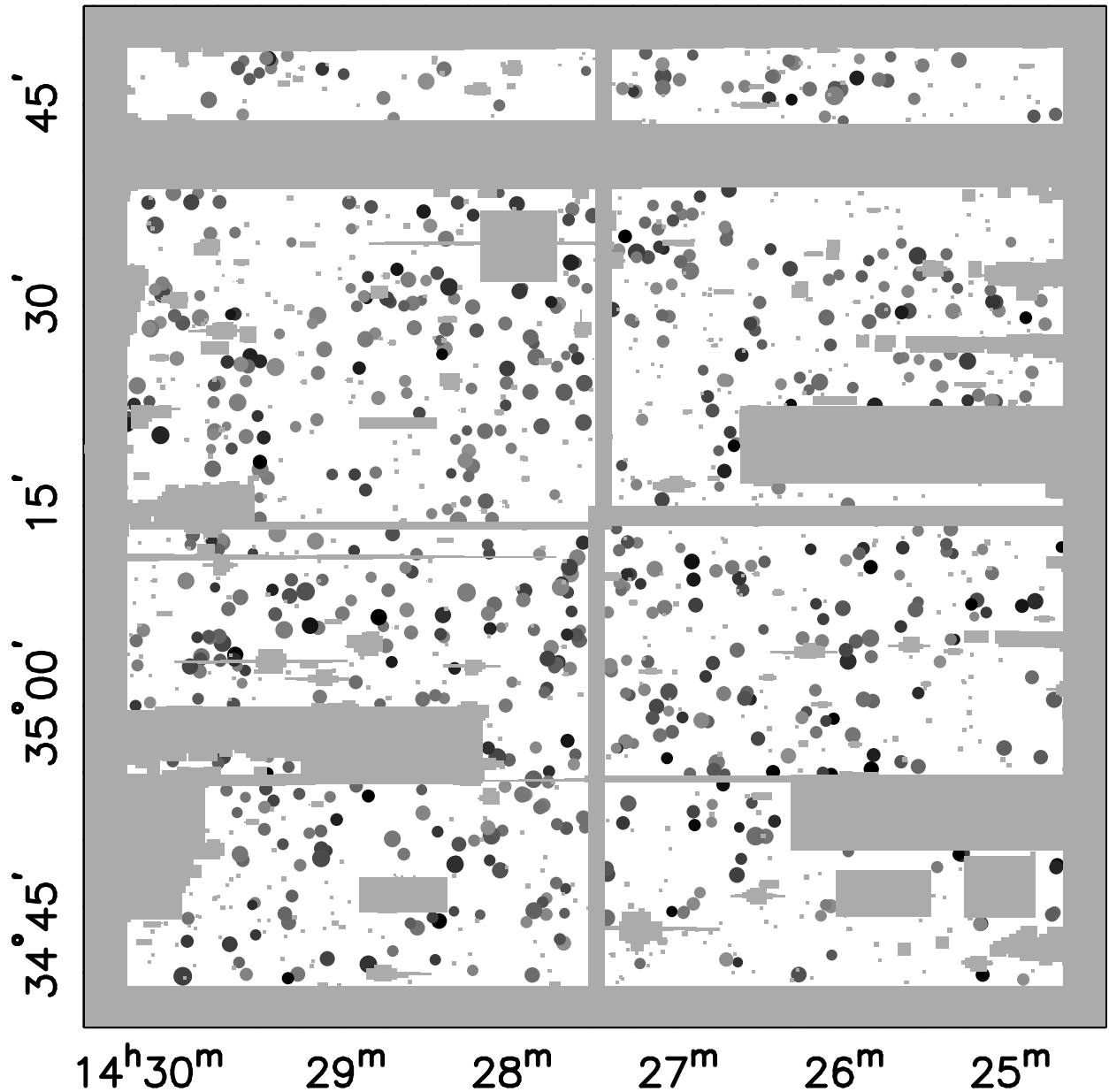}
\caption{The distribution of NDWFS EROs on the plane of the sky. 
EROs are shown with dots where the grayscale is a function of 
apparent magnitude (black dots are $K<17$) while dot size is 
inversely proportional to the photometric redshift. Masked regions
such as subfield boundaries, saturated stars, and non-photometric $K$-band data,
are shown with gray rectangles. Clustering is evident in the plot 
but there are no large gradients comparable to the sample area.
\label{fig:dist}}
\end{figure}

The final sample consists of 671 objects, of which 318 are detected in the $B_W$-band
and 635 are detected in the $R$-band. The $K<18.4$ EROs have photometric redshifts 
in the range $0.8\leq z \leq 3.0$, with the median of the distribution at $z\simeq 1.18$.
Only 5 of the 671 $K<18.4$ EROs have photometric redshifts of $z>2$.
ERO number counts as a function of $K$-band limiting magnitude are provided in 
Table~\ref{table:ang} and Figure~\ref{fig:counts}, 
along with results from previous surveys. We evaluated the uncertainties of the 
sky surface density, for the NDWFS and previous work, using the method discussed
by \cite{efs91}, which includes the contribution of large-scale structure.
The contribution of clustering to the uncertainties is typically 
several times larger than uncertainties determined by Poisson statistics.
For our $K<18.4$ ERO sample, accounting for the clustering increases the $1\sigma$ 
uncertainty from 5\% to 20\%! We note that the uncertainties quoted by some studies
do not include this contribution \citep[e.g.,][]{roc02, miy03, roc03}. 
The distribution of the ERO sample on the plane of the sky is shown in 
Figure~\ref{fig:dist}. ERO surveys of $\sim 0.1~{\rm deg}^2$ often have 
individual structures with sizes comparable to the field of view
\citep[e.g.,][]{dad00}. While 
clustering and voids are evident in Figure~\ref{fig:dist}, there are 
no obvious $\sim 0.5^\circ$ structures or gradients in the distribution of EROs 
in our sample.

\section{The Correlation Function}
\label{sec:cor}

We determined the angular correlation function using the \cite{lan93} estimator:
\begin{equation}
\hat\omega(\theta)=\frac{DD-2DR+RR}{RR}
\end{equation}
where $DD$, $DR$, and $RR$ are the number of galaxy-galaxy, galaxy-random
and random-random pairs at angular separation $\theta\pm\delta\theta/2$.
The pair counts were determined in logarithmically spaced bins between
$10^{\prime\prime}$ and $0.7^\circ$. 

We employed the same methodology as \cite{bro03} to generate random object 
catalogs, correct for the integral constraint \citep{gro77}, and estimate the
covariance of the $\hat\omega(\theta)$ bins \citep[using the technique of][]{eis01}.
The random object catalog contains 100 times the number of objects as the ERO catalog,
so $DR$ and $RR$ are renormalized accordingly. 

The angular correlation function was assumed to be a power-law given by
\begin{equation}
\omega(\theta) = \omega(1^\prime) \left( \frac{\theta}{1^\prime} \right)^{1-\gamma}
\end{equation}
where $\gamma$ is a constant. This is a good approximation of the
observed galaxy spatial correlation function from the 2dFGRS and Sloan 
Digital Sky Survey (SDSS) 
on scales of $\lesssim 10 ~h^{-1} {\rm Mpc}$ \citep{nor01,nor02,zeh02}. 
Throughout this paper we assume $\gamma=1.87$, the approximate value 
of $\gamma$ for $z<0.15$ red galaxies from the 2dFGRS and SDSS surveys.
For a $\gamma=1.87$ power-law, the integral constraint for this study 
was approximately $6\%$ of the  amplitude of the correlation function at $1^\prime$.
Pair counts and the estimate of the angular correlation function (including the integral
constraint correction) for $R-K>5.0$ and $R-K>5.5$ EROs are presented in Table~\ref{table:pairs}.

The spatial correlation function was obtained using the \cite{lim54} equation;
\begin{equation}
\omega(\theta)= \int^\infty_0  \frac{dN}{dz}
\left[ \int^\infty_0 \xi (r(\theta,z,z^\prime),z) \frac{dN}{dz^\prime} dz^\prime \right] dz
\left/
\left( \int^\infty_0 \frac{dN}{dz} dz \right)^2 \right.
\end{equation}
where $\frac{dN}{dz}$ is the redshift distribution without clustering, $\xi$ is the spatial correlation function
and $r(\theta,z,z^\prime)$ is the comoving distance between two objects at redshifts $z$
and $z^\prime$ separated by angle $\theta$ on the sky. The spatial correlation function was assumed
to be a power law given by
\begin{equation}
\xi (r,z) = [ r/r_0(z)]^{-\gamma}.
\label{eq:spa}
\end{equation}

\begin{figure}
\plottwo{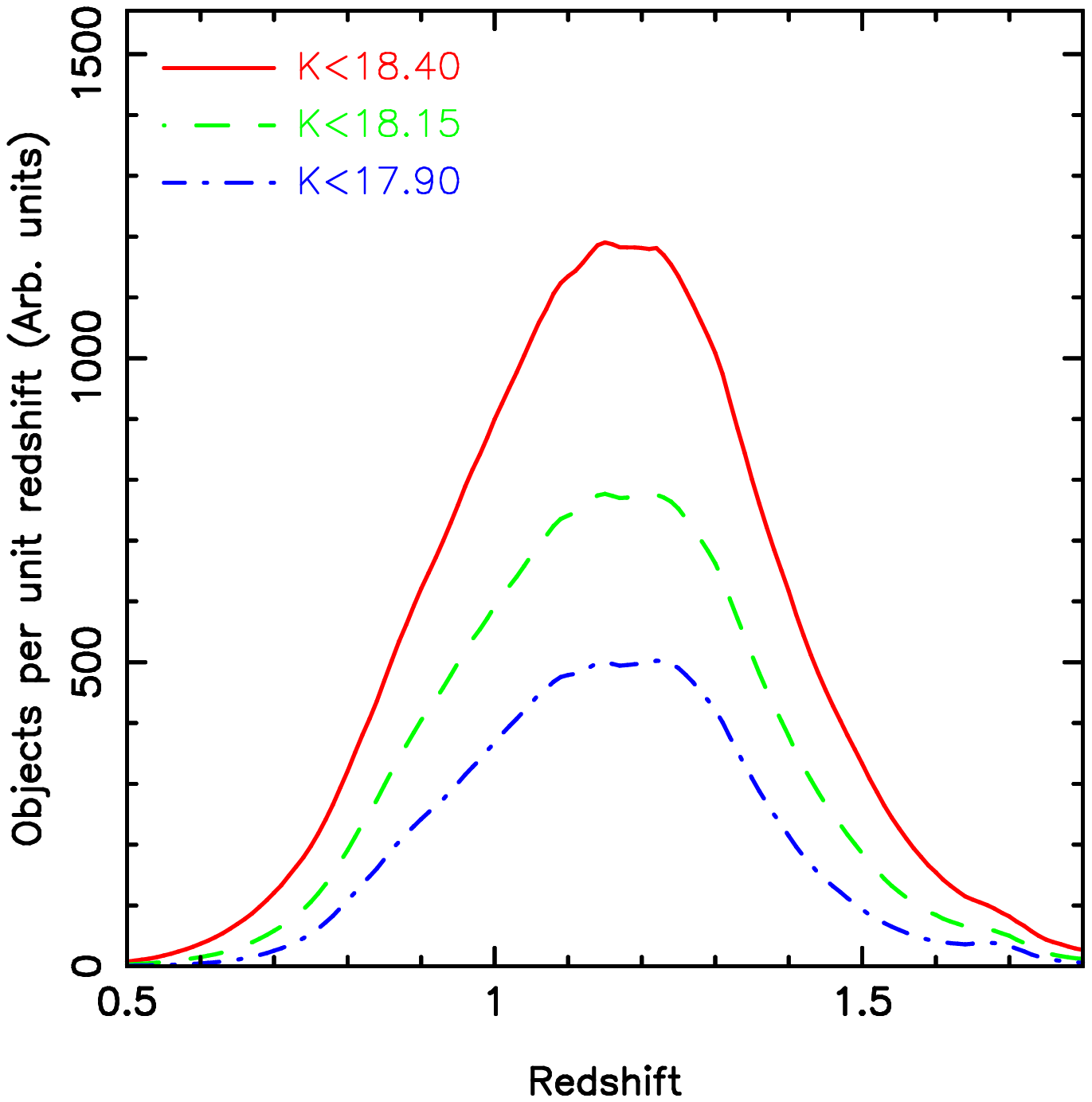}{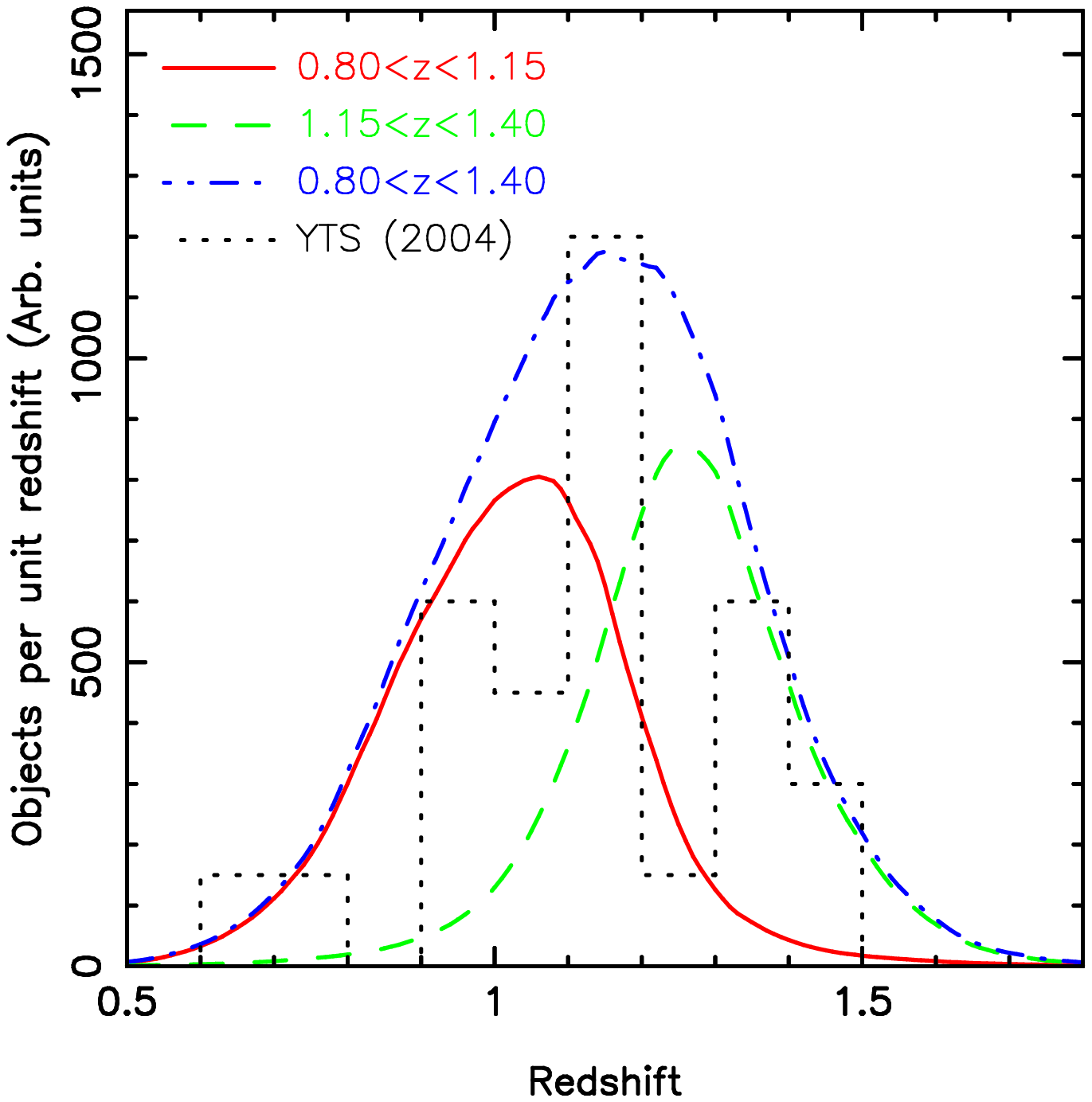}
\caption{The model redshift distributions of NDWFS EROs as a function of 
apparent magnitude (left panel) and photometric redshift
(right panel). The histogram of $K_S<18.7$ ERO spectroscopic redshifts
from \cite{yan04} is also shown.
Unlike most models of the redshift distribution of $K\gtrsim19$
EROs \citep[e.g.,][]{dad01,roc02,roc03}, the NDWFS $K<18.40$ ERO 
redshift distribution model has few objects at $z>1.5$.
\label{fig:dndz}}
\end{figure}

We estimated the redshift distribution for the sample by
summing the redshift likelihood distributions of the individual
galaxies in each subsample. Model redshift distributions for 
subsamples selected by apparent magnitude and photometric redshift are 
shown in Figure~\ref{fig:dndz}. While the individual photometric redshifts are 
not especially accurate, they do include information provided
by the observed ERO photometry and are likely to provide a fair
approximation of the ERO redshift distribution. Redshift distribution models 
which only reproduce the apparent ERO number counts and local galaxy luminosity functions
\citep[e.g.,][]{dad01,roc02,roc03} have fewer constraints and may have larger systematic errors. 
The estimated median redshift of the $K<18.40$ EROs is 1.18, 
which is almost identical to the spectroscopic median redshift of 24 
$K_S<18.7$ EROs from \cite{yan04}. The median redshift is 
also similar to $K_S<18.5$ EROs in the K20 spectroscopic sample 
\citep[][A. Cimatti 2003, private communication]{cim02}.

\section{The clustering of EROs}
\label{sec:clu}

\begin{figure}
\plotone{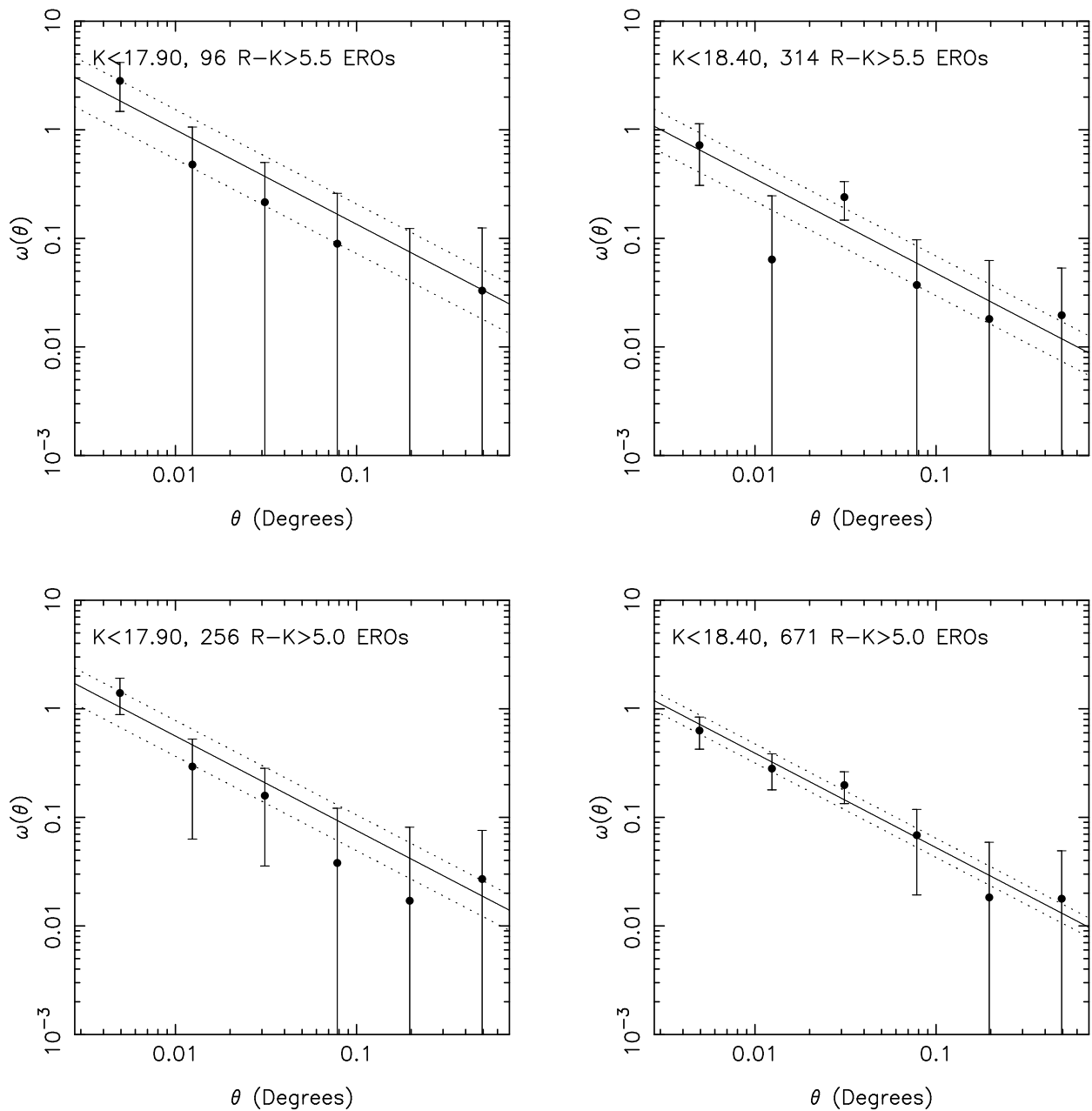}
\caption{The angular correlation function of NDWFS EROs for several apparent
$K$-band magnitude limited bins. Power-law fits to the data with $\gamma$ fixed at 
$1.87$ are shown along with $\pm 1 \sigma$ errors (dotted lines). 
\label{fig:ang}}
\end{figure}

We measured the angular and spatial correlation functions for a 
series of apparent magnitude, apparent color, absolute magnitude, and redshift bins.
A power-law of the form $\omega (\theta) =  A \theta^{1-\gamma}$ was fitted
to the data with $\gamma$ fixed to $1.87$.  Much larger imaging surveys,
including the completed NDWFS Bo\"{o}tes and Cetus fields, will have sufficient area
to accurately measure $\gamma$. When parameterizing the power-law fits, 
we use $\omega(1^\prime)$ instead of $\omega(1^\circ)$ as it depends 
less on the assumed value of $\gamma$. Using $\gamma=1.80$
instead of $\gamma=1.87$ increases $\omega(1^\prime)$ by $\simeq 10\%$
and $\omega(1^\circ)$ by $\simeq 35\%$. The best-fit values of $r_0$ do depend
on the assumed value of $\gamma$, but for the NDWFS ERO sample, changing
$\gamma$ from $1.87$ to $1.80$ increases $r_0$ by only $\simeq 10\%$. 
Measurements of $\omega(1^\prime)$ for EROs as a function of $K$-band limiting 
magnitude from our study and the literature are summarized in Table~\ref{table:ang}. 
Angular correlation functions for apparent magnitude limited samples 
are also plotted in Figure~\ref{fig:ang}.
Estimates of $\omega(1^\prime)$ and $r_0$ for each of the NDWFS 
subsamples are presented in Table~\ref{table:r0} and discussed in 
\S\ref{sec:app} to \S\ref{sec:foz}.

\subsection{Clustering as a function of apparent magnitude}
\label{sec:app}

\begin{figure}
\plotone{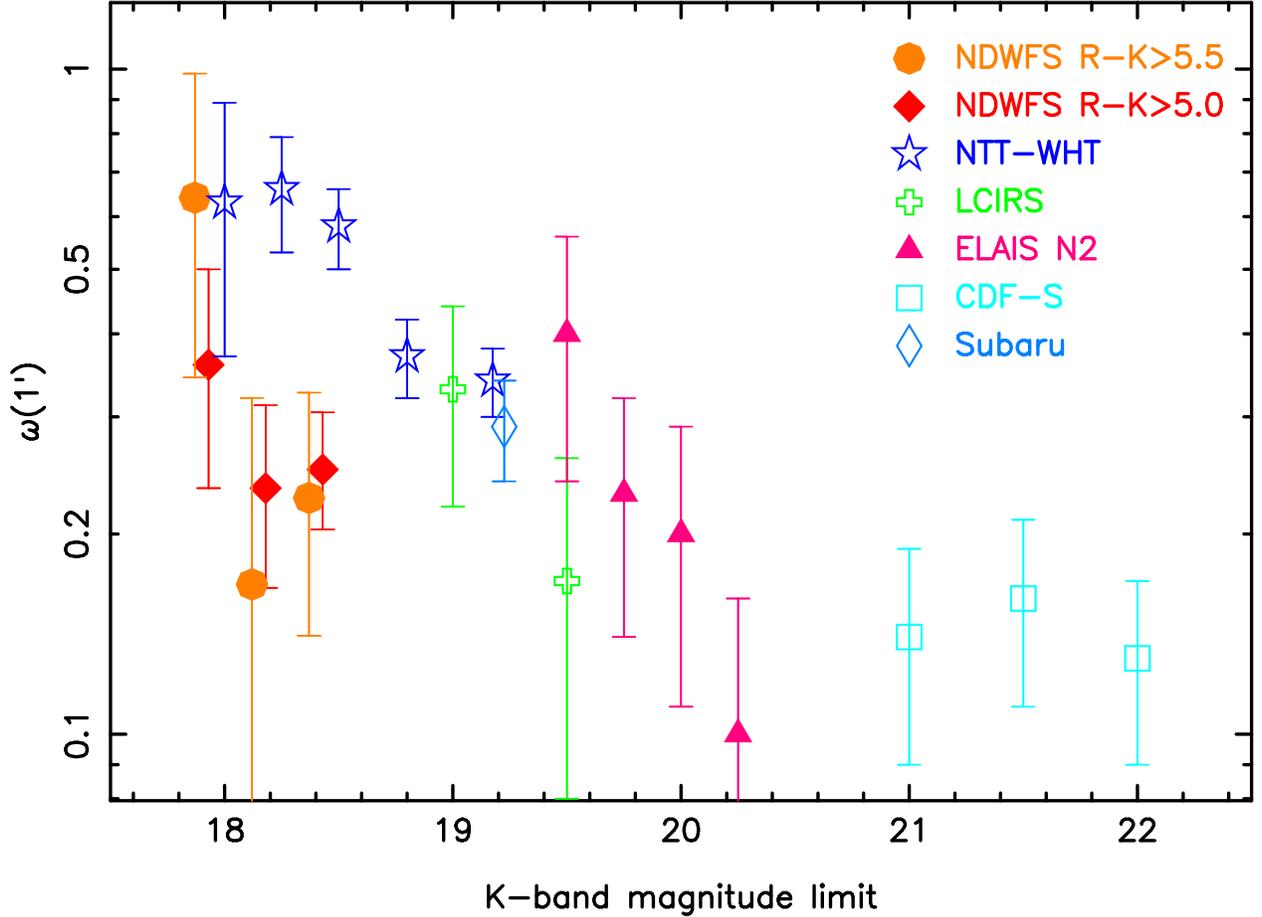}
\caption{The amplitude of the ERO angular correlation function at $1^\prime$ as a 
function of $K$-band limiting magnitude.  
The error bars show the published uncertainties, which may not include the
contribution of the covariance.
The $4~{\rm arcmin}^2$ HDF-S measurement of $K<24$ ERO clustering (not shown) 
is $\omega(1^\prime)=0.16\pm 0.10$. Combining the NDWFS with previous ERO clustering 
studies, there is evidence for the ERO angular clustering decreasing with increasing 
magnitude. Unless the redshift distribution of $K>19$ EROs is much broader than the 
redshift distribution of $K<19$ EROs, the spatial clustering of faint EROs is somewhat weaker
than the spatial clustering of $K<19$ EROs.
\label{fig:omegak}}
\end{figure}

The amplitude of the angular correlation function for a series of 
apparent magnitude limited samples is presented in Figure~\ref{fig:omegak} and 
Table~\ref{table:ang}, along with estimates from the literature.
While our $K<18.4$ sample has a larger volume and more objects than previous
studies, our uncertainties are comparable to the published uncertainties
of many previous studies. This is due to our inclusion of the covariance
when fitting a power-law to the data. Our estimates of the amplitude of 
the angular correlation function are $\sim 2\sigma$ lower than the 
smaller $K_S<18.4$ ERO samples from \cite{dad00}. Within our sample we 
do not see a significant change in the angular clustering amplitude with 
apparent magnitude, but this is not unexpected as we span a small
range of apparent magnitudes.

The first section of Table~\ref{table:r0} provides an estimate 
of $r_0$ for EROs as a function of apparent limiting magnitude. 
While the NDWFS $r_0$ values are accurate to 
$\sim 15\%$, each apparent magnitude bin spans a large range of redshift
and absolute magnitude. As $r_0$ is correlated with luminosity 
in other galaxy samples \citep[e.g.,][]{gia01,nor02,zeh02,bro03}, a correlation
between $r_0$ and apparent magnitude might be expected. We do not 
observe a significant correlation within the NDWFS, but our 
uncertainties are too large to rule out such a correlation.

Combining published ERO samples provides spatial clustering measurements
over a broad magnitude range. However, it is not possible to directly compare 
the published $r_0$ measurements of different ERO samples (Table~\ref{table:prev}),
as different authors use different models of the ERO redshift distribution.
Different studies also estimate the uncertainties of the angular correlation function
and $r_0$ using different techniques. In Table~\ref{table:pairs}, we present the NDWFS pair
counts for the $R-K>5.0$ and $R-K>5.5$ $K<18.40$ ERO angular correlation functions, so 
other researchers can apply their techniques for estimating the correlation function 
to our data. Poisson statistics underestimate the uncertainties of the 
correlation function on large-scales, where object pair counts are high and 
the uncertainties of the correlation function are dominated by large-scale 
structure. The uncertainties of clustering measurements from deep pencil-beam 
surveys should be larger than those of the NDWFS, and the large scatter of $K>19$ 
ERO clustering measurements shown in Figure~\ref{fig:omegak} may reflect this. 

If we assume the published best-fit values of the amplitude of the angular 
correlation are correct, then the angular clustering of EROs does decrease 
with increasing limiting magnitude. We find  $\omega(1^\prime)=0.25 \pm 0.05$ for 
$K<18.40$ EROs while \cite{roc03} find  $\omega(1^\prime)\sim 0.13$ for $K<22$ EROs.
Unless the redshift distribution of faint EROs is very broad, the spatial clustering
of EROs is decreasing with increasing apparent magnitude. Several studies to 
measure $r_0 \sim 10~h^{-1}{\rm Mpc}$ for faint EROs \citep{roc02,dad03,miy03,roc03},
but their model redshift distributions contain more high redshift objects than the 
GOODS $K_S<20.1$ ERO photometric redshift distribution \citep{mou04}.
If $\omega(1^\prime)\simeq 0.13$ for faint EROs and the GOODS photometric
redshifts are accurate, then $r_0$ decreases from 
$9.7 \pm 1.1 ~h^{-1} {\rm Mpc}$ for $K<18.40$ to $r_0\sim 7.5 ~h^{-1}{\rm Mpc}$
for $K\gtrsim 20$ EROs. Red galaxies at $z<1$ have a comparable range of $r_0$ values  
\citep[e.g.,][]{nor02,zeh02,bro03} and their spatial clustering is 
correlated with absolute magnitude. The current measurements of ERO clustering
are consistent with EROs being the progenitors of local red galaxies.

\subsection{Clustering as a function of apparent color}
\label{sec:color}

We present the clustering of $R-K>5.0$ and $R-K>5.5$ EROs in 
Figure~\ref{fig:omegak} and Table~\ref{table:r0}. We find the 
angular and spatial clustering of $R-K>5.5$ galaxies does not
differ significantly from the remainder of the sample.
Low redshift galaxies may have a bimodal distribution of clustering
properties as a function of color \citep{bud03}. This could be 
due to the bimodal distribution of galaxy colors at low redshift, 
or a bimodality of the clustering properties of galaxies as a function
of star formation rate. If the clustering is bimodal at all redshifts, 
we would not expect a correlation between clustering and color 
within a red galaxy sample. We do not see a correlation of 
clustering with color but a larger sample with improved photometric 
redshifts is required so accurate spatial clustering measurements 
can be performed as a function of rest-frame color. 

\subsection{Clustering as a function of absolute magnitude}

The clustering of EROs as a function of absolute magnitude is presented in 
Table~\ref{table:r0}. We have determined the absolute magnitudes (without 
evolution corrections) of the EROs using the best-fit $\tau$-model SED.
As shown in Figure~\ref{fig:photoz}, ERO photometric redshifts can have
large uncertainties and our ERO absolute magnitudes are, at best, approximations.
The two absolute magnitude bins are approximately volume limited 
samples with the same photometric redshift range. Both absolute magnitude bins
are extremely luminous, and contain EROs approximately 4 times
brighter than the local value of $L^*$ \citep[$M^*_K = -23.44\pm 0.03$;][]{col01}. 
We do not see a significant correlation between luminosity and $r_0$ within
the sample. The correlation between galaxy luminosity and clustering
is only seen unambiguously in $z<1$ samples (e.g., 2dFGRS and SDSS) which 
contain a factor of $\gtrsim 10$ more galaxies than the NDWFS ERO sample. 
A strong correlation between ERO luminosity and spatial clustering remains plausible,
and may be detected with an analysis of the complete NDWFS.

\subsection{Clustering as a function of redshift}
\label{sec:foz}

\begin{figure}
\plotone{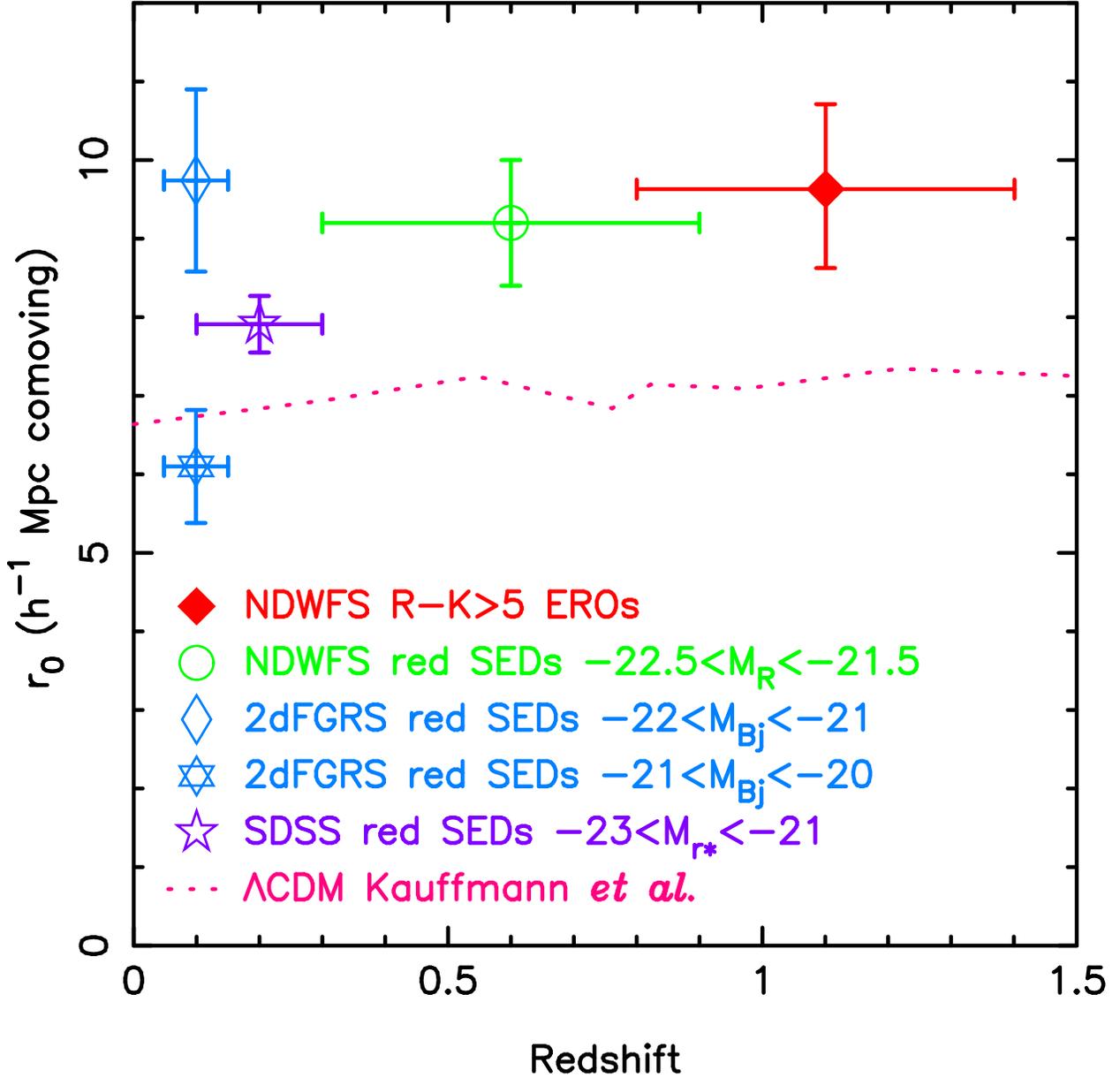}
\caption{The spatial correlation function of the most luminous red galaxies
as a function of redshift. The NDWFS ERO datapoints are shown along
with previous $r_0$ measurements of $z<1$ red galaxies from the 2dFGRS, 
SDSS and NDWFS \citep{nor02,bud03,bro03}. If $K<18.40$ EROs are the progenitors
of local $\sim 4L^*$ early-type galaxies, the data are consistent with 
little or no evolution of $r_0$ for $z<1.4$ red galaxies.
A $\Lambda$CDM model for the clustering of early-type galaxies selected by stellar mass \citep{kau99}
also exhibits little evolution, but has lower clustering than most of the samples plotted here.
\label{fig:r0z}}
\end{figure}

We measured the clustering of EROs within the sample with two 
photometric redshift bins, $0.80<z<1.15$ and $1.15<z<1.40$.
We exclude EROs beyond these redshift ranges, as they contribute
less than $10\%$ of the total $K<18.40$ ERO number counts.
The results are presented in Table~\ref{table:r0} and Figure~\ref{fig:r0z}.
We do not observe significant evolution of $r_0$ with redshift within
the ERO sample. However, our uncertainties are large and the redshift distributions
of the two samples overlap, so they are not  entirely independent.

\section{Discussion}
\label{sec:dis}

The clustering of $0.80<z<1.40$, $K<18.40$ EROs is well approximated by a
power law with $r_0=9.6\pm 1.0 ~h^{-1} {\rm Mpc}$ and $\gamma$ fixed at $1.87$. 
As EROs are thought to be the progenitors of local ellipticals, it is useful
to compare the clustering measurements of these populations.
In the 2dFGRS, the spatial correlation function increases from 
$r_0=6.10\pm0.72 ~h^{-1} {\rm Mpc}$ for $\simeq 2L^*$ red galaxies
to $r_0=9.74\pm1.16~h^{-1} {\rm Mpc}$ for $\simeq 4L^*$ red galaxies \citep{nor02}.
Other $z<1$ surveys, including the NDWFS, measure comparable spatial 
clustering for red galaxies \citep[e.g.,][]{wil98,bud03,bro03}.
It is not unreasonable to assume the brightest EROs are the progenitors 
of the $\simeq 4L^*$ red galaxies in the local Universe, as the comoving 
spatial clustering of the two populations is comparable. However, this
assumes models predicting little or no evolution of the $\gtrsim L^*$ 
galaxy correlation function \citep[e.g.,][]{kau99,ben01} are valid.

If $K<18.40$ EROs are the progenitors of the most luminous local
red galaxies, fainter EROs could the progenitors of $\sim L^*$
red galaxies. Several previous studies of fainter EROs find
they are very strongly clustered, with $r_0\sim 10~h^{-1} {\rm Mpc}$
\citep{roc02,dad03,miy03,roc03}. This is much stronger than the clustering
of local $L^*$ red galaxies, where $r_0 \simeq 6 h^{-1}{\rm Mpc}$ \citep{nor02,zeh02}. However, the 
ERO spatial clustering measurements could be subject to large, and possibly systematic, errors.
Several of these measurements use model redshift distributions
which are primarily constrained by the local galaxy luminosity function
and faint galaxy number counts \citep{dad01,roc02,roc03}. 
Other model redshift distributions use unverified photometric redshifts
\citep{miy03} or redshifts which could only be verified with galaxies other than
EROs \citep{fir02}. As shown in Figure~\ref{fig:omegak}, the angular clustering of EROs does decrease with increasing
apparent magnitude. Unless the redshift distribution of $K \gtrsim 20$ EROs is 
very broad, the spatial clustering of $K \gtrsim 20$ EROs is weaker than
the spatial clustering of $K < 18.40$ EROs. 

While $K<18.40$ EROs may be the progenitors of most luminous local ellipticals, 
our spatial clustering measurement should be treated with some caution.
Our ERO sample spans broad ranges of redshift and absolute magnitude
($-27.9\lesssim M_K - 5 {\rm log} h \lesssim -23.9$). 
The luminosity and density evolution of EROs and red galaxies at $z<1$
has not been accurately determined. The PEGASE $\tau$ models predict 
$\sim 0.8$ magnitudes of luminosity evolution at $z<1$, so EROs would be 
the progenitors to  $\simeq 2L^*$ red galaxies. If this were the case,
the spatial correlation function would be decreasing with decreasing redshift, 
which is unphysical. The uncertain luminosity and density evolution of EROs 
limits the use of $>L^*$ EROs to measure the evolution of the galaxy spatial 
correlation function.

Current ERO spatial clustering measurements, including our study, have 
large uncertainties and may be subject to systematic errors.
We will significantly reduce the random uncertainties of ERO clustering
measurements when we analyze the entire NDWFS Bo\"{o}tes field.
The uncertainties and systematic errors of our photometric redshifts
will be accurately determined as we obtain more spectroscopic
redshifts. We will also improve the photometric
redshifts for EROs by using NDWFS, FLAMINGOS \citep{gon04}, and {\it Spitzer} Space Telescope 
data. We will then be able to accurately measure 
the spatial clustering of EROs as a function of luminosity, color,
and redshift.

\section{Summary}
\label{sec:sum}

We have measured the clustering of 671 $K<18.40$ EROs with $0.98~{\rm deg}^2$ subset of the 
NDWFS. This study covers an area nearly 5 times larger and has twice the 
sample size of any previous ERO clustering study.
The angular clustering of $K<18.40$, $R-K>5.0$ EROs is well described by 
a power-law with  $\omega(1^\prime)=0.25 \pm 0.05$ and $\gamma=1.87$.
Using a model of the ERO redshift distribution derived from photometric redshifts,
we find the spatial clustering of $K<18.40$ EROs is given by 
$r_0=9.7 \pm 1.0 ~h^{-1} {\rm Mpc}$ comoving. Within our study, we  
detect no significant correlations between ERO clustering and apparent magnitude, 
apparent color, absolute magnitude, or redshift. 
However, our uncertainties are large and such correlations may exist.
When combined with data from other studies, there is evidence of the angular
clustering of EROs decreasing with increasing apparent magnitude. 
Unless the redshift distribution of $K \gtrsim 20$ EROs is very broad, the spatial
clustering of EROs decreases with increasing apparent magnitude.
As the uncertainties and systematic errors of current ERO spatial clustering 
measurements are large, they do not yet provide strong tests of models of 
structure evolution and galaxy formation. 

\acknowledgments

This research was supported by the National Optical Astronomy Observatory which is
operated by the Association of Universities for Research in Astronomy (AURA), Inc.
under a cooperative agreement with the National Science Foundation.
We thank our colleagues on the NDWFS team and the KPNO and CTIO observing support staff.
We are grateful to Frank Valdes, Lindsey Davis and the IRAF team for the packages used to reduce the 
imaging data. We thank Alyson Ford, Chris Greer, Heather Gross, Lissa Miller, and Erin Ryan, for
helping reduce KPNO Mosaic-1 and ONIS imaging data used for our work.
While working on this paper, the authors had many useful discussions about galaxy 
clustering with Tod Lauer and Daniel Eisenstein.

\clearpage

\clearpage


\begin{deluxetable}{ccccccccccc}
\rotate
\tabletypesize{\scriptsize}
\tablecaption{A summary of ERO spatial clustering studies.
\label{table:prev}}
\tablewidth{0pt}
\tablehead{
\colhead{Survey\tablenotemark{a}} &
\colhead{Area} &
\colhead{Number of} &
\colhead{Magnitude} &
\colhead{Selection} &
\colhead{Additional} &
\colhead{Measured or} &
\colhead{$z$ distribution\tablenotemark{b}} &
\colhead{$r_0$ comoving\tablenotemark{c}} &
\colhead{Assumed} \\
\colhead{} &
\colhead{(${\rm arcmin}^2$)} &
\colhead{Galaxies} &
\colhead{Range} &
\colhead{} &
\colhead{selection criteria} &
\colhead{model $z$ range} &
\colhead{model} &
\colhead{$(h^{-1} {\rm Mpc})$} &
\colhead{value of $\gamma$\tablenotemark{d}} 
}
\startdata
NDWFS    & 3529 & 671  & $K <18.40$ & $R-K>5$   & & $0.8\lesssim z\lesssim 3.0$               & PhotZ   & $9.7\pm 1.0$          & 1.87\\ 
\\
K20      & 52   & 18   & $K <19.2$  & $R-K>5$   & Dusty SF SED    & $0.796 \leq z \leq 1.419$ & Spectra & $\lesssim 2.5$        & 1.8 \\ 
K20      & 52   & 15   & $K <19.2$  & $R-K>5$   & Old stellar SED & $0.726 \leq z \leq 1.222$ & Spectra & 5.5 to 16             & 1.8 \\ 
NTT-WHT  & 701  & 400  & $K <19.2$  & $R-K>5$   & & $0.8\lesssim z\lesssim 2.0$               & PE      & $13.8\pm 1.5$         & 1.8 \\ 
LCIRS    & 744  & 337  & $H <20.0$  & $R-H>4$   & & $0.7\lesssim z\lesssim 1.5$               & PhotZ   & $11.1\pm 2.0$         & 1.8 \\ 
LCIRS    & 407  & 312  & $H <20.5$  & $R-H>4$   & & $0.7\lesssim z\lesssim 1.5$               & PhotZ   & $7.7\pm 2.4$          & 1.8 \\
Subaru   & 114  & 134  & $K <20.2$  & $R-K_S>5$ & Dusty SF SED    & $0.0<z<4.3$               & PhotZ   & $12\pm 2$             & 1.8 \\ 
Subaru   & 114  & 143  & $K <20.2$  & $R-K_S>5$ & Old stellar SED & $0.0<z<4.3$               & PhotZ   & $11\pm 1$             & 1.8 \\ 
ELAIS N2 & 81.5 & 158  & $K <20.25$ & $R-K>5$   & & $1\lesssim z\lesssim 3$                   & M-DE    & $12.8\pm1.5$          & 1.8 \\ 
ELAIS N2 & 81.5 & 158  & $K <20.25$ & $R-K>5$   & & $1\lesssim z\lesssim 3$                   & NE      & $10.3\pm1.2$          & 1.8 \\ 
CDF-S    & 50.4 & 198  & $K_S<22.0$ & $I_{775}-K_S>3.92$ & & $1\lesssim z\lesssim 3$          & M-DE    & $12.5\pm1.2$          & 1.8 \\ 
HDF-S    & 4    & 18   & $K <24.0$  & $I-K>4$   & &                                           & PhotZ   & $16.9\pm^{2.9}_{5.5}$ & 1.8 \\ 
HDF-S    & 4    & 39   & $K <24.0$  & $I-K>3.5$ & &                                           & PhotZ   & $6.2\pm^{5.4}_{7.1}$  & 1.8 \\ 
HDF-S    & 4    & 23   & $K <24.0$  & $I-K>3.5$ & $0.8<z<2.0$ & $0.8\lesssim z\lesssim 2.0$   & PhotZ   & $9.7\pm 2.0$          & 1.8 \\ 
\enddata 
\tablenotetext{a}{CDF-S \citep{roc03}, ELAIS N2 \citep{roc02}, HDF-S \citep{dad03}, NTT-WHT \citep{dad01}, 
K20 \citep{dad02}, LCIRS \citep{fir02}, Subaru \citep{miy03}.}
\tablenotetext{b}{M-DE \citep[merging and density evolution;][]{roc02}, NE \citep[no evolution;][]{roc02}, 
PE \citep[single burst and passive evolution;][]{dad01}, PhotZ \citep[photometric redshifts;][; this work]{fir02,dad03}}
\tablenotetext{c}{Values of $r_0$ are for a $\Omega_m=0.3$, $\Lambda=0.7$ cosmology. Uncertainties are as published,
and were determined using a variety of techniques.}
\tablenotetext{d}{For this study, changing the value of $\gamma$ from 1.87 to 1.80 increases $r_0$ by $\simeq 10\%$.}
\end{deluxetable}

\begin{deluxetable}{cccccccc}
\tablecolumns{8}
\tabletypesize{\scriptsize}
\tablecaption{A summary of ERO angular clustering studies including number counts and
sky surface density.
\label{table:ang}}
\tablewidth{0pt}
\tablehead{
\colhead{Survey\tablenotemark{a}} &
\colhead{Area} &
\colhead{Number} &
\colhead{of EROs} &
\colhead{Magnitude} &
\colhead{Selection} &
\colhead{$\omega(1^\prime)$\tablenotemark{c}} &
\colhead{Assumed} \\
\colhead{} &
\colhead{(${\rm arcmin}^2$)} &
\colhead{EROs} &
\colhead{per ${\rm deg}^2$\tablenotemark{b}} &
\colhead{Range} &
\colhead{} &
\colhead{} &
\colhead{ $\gamma$ value} 
}
\startdata
\cutinhead{This study}
NDWFS    & 3529   & 256  & $2.6\pm0.4\times10^2$ & $K<17.90$   & $R-K>5.0$      & $0.36\pm 0.13$   & 1.87 \\ 
NDWFS    & 3529   & 421  & $4.3\pm0.6\times10^2$ & $K<18.15$   & $R-K>5.0$      & $0.23\pm 0.07$   & 1.87 \\ 
NDWFS    & 3529   & 671  & $6.8\pm0.9\times10^2$ & $K<18.40$   & $R-K>5.0$      & $0.25\pm 0.05$   & 1.87 \\ 
\cutinhead{Previous studes ordered by limiting magnitude}
NTT-WHT  & 701    & 58   & $2.9\pm1.0\times10^2$ & $K_S<18.00$ & $R-K_S>5.0$    & $0.63\pm 0.26$   & 1.8  \\ 
NTT-WHT  & 701    & 106  & $5.4\pm1.8\times10^2$ & $K_S<18.25$ & $R-K_S>5.0$    & $0.66\pm 0.13$   & 1.8  \\ 
NTT-WHT  & 701    & 158  & $8.1\pm2.5\times10^2$ & $K_S<18.40$ & $R-K_S>5.0$    & $0.58\pm 0.08$   & 1.8  \\ 
NTT-WHT  & 701    & 279  & $1.4\pm0.4\times10^3$ & $K_S<18.80$ & $R-K_S>5.0$    & $0.37\pm 0.05$   & 1.8  \\ 
LCIRS    & 744    & 337  & $1.6\pm0.4\times10^3$ & $H<20.0$    & $R-H>4.0$      & $0.33\pm 0.11$   & 1.8  \\ 
LCIRS    & 744    & 201  & $9.7\pm2.4\times10^2$ & $H<20.0$    & $I-H>3.0$      & $0.36\pm 0.18$   & 1.8  \\ 
NTT-WHT  & 447.5  & 281  & $2.3\pm0.6\times10^3$ & $K_S<19.20$ & $R-K_S>5.0$    & $0.34\pm 0.04$   & 1.8  \\ 
Subaru   & 114    & 111  & $3.5\pm1.1\times10^3$ & $K_S<19.2$  & $R-K_S>5.0$    & $0.29\pm 0.05$   & 1.8  \\ 
LCIRS    & 407    & 312  & $2.8\pm0.5\times10^3$ & $H<20.5$    & $R-H>4.0$      & $0.17\pm 0.09$   & 1.8  \\ 
LCIRS    & 407    & 170  & $1.5\pm0.3\times10^3$ & $H<20.5$    & $I-H>3.0$      & $0.20\pm 0.16$   & 1.8  \\ 
ELAIS N2 & 81.5   & 73   & $3.2\pm1.3\times10^3$ & $K<19.50$   & $R-K>5.0$          & $0.40\pm 0.16$ & 1.8 \\ 
ELAIS N2 & 81.5   & 93   & $4.1\pm1.3\times10^3$ & $K<19.75$   & $R-K>5.0$          & $0.23\pm 0.09$ & 1.8 \\ 
ELAIS N2 & 81.5   & 112  & $4.9\pm1.4\times10^3$ & $K<20.00$   & $R-K>5.0$          & $0.20\pm 0.09$ & 1.8 \\ 
ELAIS N2 & 38.7   & 63   & $5.9\pm1.5\times10^3$ & $K<20.25$   & $R-K>5.0$          & $0.10\pm 0.06$ & 1.8 \\ 
CDF-S    & 50.4  & 137   & $9.8\pm2.6\times10^3$ & $K_S<21.00$ & $I_{775}-K_S>3.92$ & $0.14\pm 0.05$ & 1.8 \\ 
CDF-S    & 50.4  & 179   & $1.3\pm0.4\times10^4$ & $K_S<21.50$ & $I_{775}-K_S>3.92$ & $0.16\pm 0.05$ & 1.8 \\ 
CDF-S    & 50.4  & 198   & $1.4\pm0.4\times10^4$ & $K_S<22.00$ & $I_{775}-K_S>3.92$ & $0.13\pm 0.04$ & 1.8 \\ 
HDF-S    & 4     & 18    & $1.6\pm0.8\times10^4$ & $K<24.00$   & $I-K>4$            & $0.16\pm 0.10$ & 1.8 \\ 
\enddata
\tablenotetext{a}{CDF-S \citep{roc03}, ELAIS N2 \citep{roc02}, HDF-S \citep{dad03}, NTT-WHT \citep{dad00}, LCIRS \citep{fir02}, Subaru \citep{miy03}.}
\tablenotetext{b}{The sky surface density has not corrected for the contribution of Malmquist bias. 
$1\sigma$ uncertainties assume Gaussian errors and include the contribution of the integral constraint
\citep[using the methodology of][]{efs91}.}
\tablenotetext{c}{Uncertainties for $\omega(1^\prime)$ are as published and may not include the effect of the covariance on the uncertainty estimates.}
\end{deluxetable}

\begin{deluxetable}{cccccc}
\tabletypesize{\scriptsize}
\tablecaption{The angular correlation functions of $K<18.40$ EROs from $0.98~{\rm deg}^2$ of the NDWFS.
\label{table:pairs}}
\tablehead{
\colhead{Color selection} &
\colhead{Angular Scales} &
\colhead{$\omega(\theta)$} &
\colhead{$DD$} &
\colhead{$DR\times 10^2$} &
\colhead{$RR\times 10^4$}}
\startdata
$R-K>5.0$ & $0.0028^\circ$ to $0.0070^\circ$ & $0.631\pm0.204$ &         88 &       5515 &     558532 \\
$R-K>5.0$ & $0.0070^\circ$ to $0.0175^\circ$ & $0.282\pm0.100$ &        424 &      33602 &    3346550 \\
$R-K>5.0$ & $0.0175^\circ$ to $0.0440^\circ$ & $0.198\pm0.062$ &       2274 &     192715 &   19207198 \\
$R-K>5.0$ & $0.0440^\circ$ to $0.1106^\circ$ & $0.069\pm0.046$ &      12084 &    1115798 &  107760586 \\
$R-K>5.0$ & $0.1106^\circ$ to $0.2778^\circ$ & $0.018\pm0.037$ &      60806 &    5811770 &  555035500 \\
$R-K>5.0$ & $0.2778^\circ$ to $0.6977^\circ$ & $0.018\pm0.028$ &     244670 &   23592613 & 2273794800 \\
\\
$R-K>5.5$ & $0.0028^\circ$ to $0.0070^\circ$ & $0.722\pm0.409$ &         20 &       1193 &     122608 \\
$R-K>5.5$ & $0.0070^\circ$ to $0.0175^\circ$ & $0.064\pm0.177$ &         80 &       7502 &     733626 \\
$R-K>5.5$ & $0.0175^\circ$ to $0.0440^\circ$ & $0.240\pm0.087$ &        510 &      41951 &    4199066 \\
$R-K>5.5$ & $0.0440^\circ$ to $0.1106^\circ$ & $0.037\pm0.054$ &       2546 &     242766 &   23547906 \\
$R-K>5.5$ & $0.1106^\circ$ to $0.2778^\circ$ & $0.018\pm0.038$ &      13240 &    1269134 &  121503332 \\
$R-K>5.5$ & $0.2778^\circ$ to $0.6977^\circ$ & $0.020\pm0.028$ &      53148 &    5137427 &  497077474 \\
\\
\enddata
\end{deluxetable}

\begin{deluxetable}{cccccccc}
\tablecolumns{8}
\tabletypesize{\scriptsize}
\tablecaption{The angular and spatial correlation functions of EROs from $0.98~{\rm deg}^2$ of the NDWFS.
\label{table:r0}}
\tablewidth{0pt}
\tablehead{
\colhead{Selection} &
\colhead{Photometric} &
\colhead{Absolute} &
\colhead{Apparent} &
\colhead{Number} &
\colhead{$\omega(1^\prime)$} &
\colhead{Median} &
\colhead{$r_0$}\\
\colhead{criterion} &
\colhead{$z$ range} &
\colhead{magnitude range} &
\colhead{magnitude range} &
\colhead{of EROs} &
\colhead{} &
\colhead{$z$} &
\colhead{$(h^{-1} {\rm Mpc})$}
}
\startdata
\cutinhead{$R-K>5.0$ EROs selected by apparent magnitude}
$R-K>5.0$ &  0.80-3.00 & $-27.91<M_K<-24.29$ & $15.92\leq K\leq 17.90$ &  256 & $0.36 \pm 0.13$ & 1.17 & $11.0 \pm 2.2$ \\ 
$R-K>5.0$ &  0.80-3.00 & $-27.91<M_K<-23.95$ & $15.92\leq K\leq 18.15$ &  421 & $0.23 \pm 0.07$ & 1.17 & $9.1 \pm 1.5$ \\  
$R-K>5.0$ &  0.80-3.00 & $-27.91<M_K<-23.79$ & $15.92\leq K\leq 18.40$ &  671 & $0.25 \pm 0.05$ & 1.18 & $9.7 \pm 1.1$ \\  
\cutinhead{$R-K>5.5$ EROs selected by apparent magnitude}
$R-K>5.5$ &  0.80-3.00 & $-27.91<M_K<-24.43$ & $15.92\leq K\leq 17.90$ &   96 & $0.64 \pm 0.32$ & 1.22 & $13.6 \pm 3.7$ \\ 
$R-K>5.5$ &  0.80-3.00 & $-27.91<M_K<-24.25$ & $15.92\leq K\leq 18.15$ &  180 & $0.17 \pm 0.13$ & 1.22 & $7.0 \pm 3.1$ \\  
$R-K>5.5$ &  0.80-3.00 & $-27.91<M_K<-24.11$ & $15.92\leq K\leq 18.40$ &  314 & $0.23 \pm 0.09$ & 1.24 & $8.6 \pm 1.9$ \\  
\cutinhead{$R-K>5.0$ EROs selected by absolute magnitude}
$R-K>5.0$ &  0.80-1.25 & $-26.00<M_K<-25.00$ & $16.77\leq K\leq 18.40$ &  108 & $0.59 \pm 0.29$ & 1.16 & $11.4 \pm 3.0$ \\ 
$R-K>5.0$ &  0.80-1.25 & $-25.00<M_K<-24.50$ & $17.35\leq K\leq 18.40$ &  208 & $0.28 \pm 0.12$ & 1.11 & $9.0 \pm 2.1$ \\  
\cutinhead{$R-K>5.0$ EROs selected by photometric redshift} 
$R-K>5.0$ &  0.80-1.15 & $-26.44<M_K<-23.79$ & $16.28\leq K\leq 18.40$ &  318 & $0.40 \pm 0.10$ & 1.04 & $10.3 \pm 1.4$ \\ 
$R-K>5.0$ &  1.15-1.40 & $-26.69<M_K<-24.46$ & $16.27\leq K\leq 18.40$ &  292 & $0.35 \pm 0.10$ & 1.28 & $9.2 \pm 1.4$ \\  
$R-K>5.0$ &  0.80-1.40 & $-26.69<M_K<-23.79$ & $16.27\leq K\leq 18.40$ &  610 & $0.27 \pm 0.06$ & 1.15 & $9.6 \pm 1.0$ \\  
\enddata
\end{deluxetable}


\begin{thebibliography}{}

\bibitem[Alexander et al.(2002)]{ale02} Alexander, D.~M., 
Vignali, C., Bauer, F.~E., Brandt, W.~N., Hornschemeier, A.~E., Garmire, 
G.~P., \& Schneider, D.~P.\ 2002, \aj, 123, 1149 

\bibitem[Benson et al.(2001)]{ben01}
Benson, A. J., Frenk, C. S., Baugh, C. M., Cole, S., \& 
Lacey, C. G., 2001, \mnras, 327, 1041

\bibitem[Bertin \& Arnouts(1996)]{ber96}
Bertin, E., \& Arnouts, S. 1996, A\&AS, 117, 393

\bibitem[Brown et al.(2003)]{bro03} Brown, M.~J.~I., Dey, A., 
Jannuzi, B.~T., Lauer, T.~R., Tiede, G.~P., \& Mikles, V.~J.\ 2003, \apj, 
597, 225 

\bibitem[Budav{\' a}ri et al.(2003)]{bud03} Budav{\' a}ri, 
T.~et al.\ 2003, \apj, 595, 59 

\bibitem[Cimatti et al.(2002)]{cim02} 
Cimatti, A.~et al.\ 2002, \aap, 381, L68 

\bibitem[Cole et al.(2001)]{col01} 
Cole, S.~et al.\ 2001, \mnras, 326, 255 

\bibitem[Daddi et al.(2000)]{dad00} Daddi, E., Cimatti, A., 
Pozzetti, L., Hoekstra, H., R{\" o}ttgering, H.~J.~A., Renzini, A., 
Zamorani, G., \& Mannucci, F.\ 2000, \aap, 361, 535 

\bibitem[Daddi et al.(2001)]{dad01}
Daddi, E., Broadhurst, T., Zamorani, G., Cimatti, A., 
R$\ddot{\rm o}$ttgering, H., \& Renzini, A., 2001, \aap, 376, 825

\bibitem[Daddi et al.(2002)]{dad02} Daddi, E.~et al.\ 2002, 
\aap, 384, L1 

\bibitem[Daddi et al.(2003)]{dad03} 
Daddi, E.~et al.\ 2003, \apj, 588, 50 

\bibitem[Devriendt, Guiderdoni, \& Sadat(1999)]{dev99} 
Devriendt, J.~E.~G., Guiderdoni, B., \& Sadat, R.\ 1999, \aap, 350, 381 

\bibitem[Dey, Spinrad, \& Dickinson(1995)]{dey95} Dey, A., 
Spinrad, H., \& Dickinson, M.\ 1995, \apj, 440, 515 

\bibitem[Dey et al.(1999)]{dey99} Dey, A., Graham, J.~R., 
Ivison, R.~J., Smail, I., Wright, G.~S., \& Liu, M.~C.\ 1999, \apj, 519, 
610 

\bibitem[Donas, Milliard, \& Laget(1995)]{don95} 
Donas, J., Milliard, B., \& Laget, M.\ 1995, \aap, 303, 661 

\bibitem[Eisenstein \& Zaldarriaga(2001)]{eis01} 
Eisenstein, D.~J.~\& Zaldarriaga, M.\ 2001, \apj, 546, 2 

\bibitem[Efstathiou et al.(1991)]{efs91}
Efstathiou, G., Bernstein, G., Tyson, J. A., 
Katz, N., \& Guhathakurta, P., 1991, \apj, 380, L47

\bibitem[Elston, Rieke, \& Rieke(1988)]{els88} 
Elston, R., Rieke, G.~H., \& Rieke, M.~J.\ 1988, \apjl, 331, L77 

\bibitem[Falco et al.(1999)]{fal99}
Falco, E. E., et al., 1999, \apj, 523, 617

\bibitem[Fioc \& Rocca-Volmerange(1997)]{fio97}
Fioc, M.~\& Rocca-Volmerange, B.\ 1997, \aap, 326, 950

\bibitem[Firth et al.(2002)]{fir02}
Firth, A. E., et al., 2002, \mnras, 332, 617

\bibitem[Giavalisco \& Dickinson(2001)]{gia01} 
Giavalisco, M.~\& Dickinson, M.\ 2001, \apj, 550, 177 

\bibitem[Gonzalez et al.(2004)]{gon04} Gonzalez, A., et al.\ 
2004, American Astronomical Society Meeting, 204,  

\bibitem[Groth \& Peebles(1977)]{gro77}
Groth, E. J., \& Peebles, P. J. E., 1977, \apj, 217, 385

\bibitem[Hu \& Ridgway(1994)]{hu94} Hu, E.~M.~\& Ridgway, 
S.~E.\ 1994, \aj, 107, 1303 

\bibitem[Jannuzi \& Dey(1999)]{jan99}
Jannuzi, B. T., \& Dey, A., 1999, in ASP Conf. Ser. 191, 
Photometric Redshifts and High Redshift Galaxies,
 ed. R. J. Weymann, L. J. Storrie-Lombardi, M. Sawicki, \& R. J. Brunner 
(San Francisco: ASP), 111

\bibitem[Kauffmann et al.(1999)]{kau99}
Kauffmann, G., Colberg, J. M., Diaferio, A., \& 
White, S. D. ., 1999, \mnras, 307, 529

\bibitem[Kron(1980)]{kro80}
Kron, R. G., 1980, \apjs, 43, 305

\bibitem[Landy \& Szalay(1993)]{lan93}
Landy, S. D., \& Szalay, A. S. 1993, \apj, 412, 64

\bibitem[Limber (1954)]{lim54}
Limber, N. D., \apj, 119, 655


\bibitem[Madgwick et al.(2002)]{mad02}
Madgwick, D. S., et al. 2002, \mnras, 332, 827

\bibitem[Malmquist (1920)]{mal20}
Malmquist, K.G. 1920, Lund Medd. Ser. II, 22, 1

\bibitem[McCarthy, Persson, \& West(1992)]{mcc92} McCarthy, 
P.~J., Persson, S.~E., \& West, S.~C.\ 1992, \apj, 386, 52 

\bibitem[Miyazaki et al.(2003)]{miy03} Miyazaki, M., et al.\ 
2003, \pasj, 55, 1079 

\bibitem[Mobasher et al.(2004)]{mob04} Mobasher, B.~et al.\ 
2004, \apjl, 600, L167 

\bibitem[Moriondo, Cimatti, \& Daddi(2000)]{mor00} Moriondo, 
G., Cimatti, A., \& Daddi, E.\ 2000, \aap, 364, 26 

\bibitem[Moustakas et al.(2004)]{mou04} 
Moustakas, L.~A.~et al.\ 2004, \apjl, 600, L131 

\bibitem[Norberg et al.(2001)]{nor01}
Norberg, P., et al. 2001, \mnras, 328, 64

\bibitem[Norberg et al.(2002)]{nor02}
Norberg, P., et al. 2002, \mnras, 332, 827

\bibitem[Roche et al.(2002)]{roc02} Roche, N.~D., Almaini, 
O., Dunlop, J., Ivison, R.~J., \& Willott, C.~J.\ 2002, \mnras, 337, 1282 

\bibitem[Roche, Dunlop, \& Almaini(2003)]{roc03} 
Roche, N.~D., Dunlop, J., \& Almaini, O.\ 2003, \mnras, 346, 803 

\bibitem[Somerville et al.(2001)]{som01}
Somerville, R. S., Lemson, G., Sigad, Y., 
Dekel, A., Kauffmann, G., \& White, S. D. M., 2001, \mnras, 320, 289

\bibitem[Somerville et al.(2004)]{som04} Somerville, R.~S., 
Lee, K., Ferguson, H.~C., Gardner, J.~P., Moustakas, L.~A., \& Giavalisco, 
M.\ 2004, \apjl, 600, L171 

\bibitem[Spinrad et al.(1997)]{spi97} Spinrad, H., Dey, A., 
Stern, D., Dunlop, J., Peacock, J., Jimenez, R., \& Windhorst, R.\ 1997, 
\apj, 484, 581 

\bibitem[Stiavelli \& Treu(2001)]{sti01} 
Stiavelli, M.~\& Treu, T.\ 2001, ASP Conf.~Ser.~230: Galaxy Disks and Disk Galaxies, 603 

\bibitem[Willmer, da Costa, \& Pellegrini(1998)]{wil98} 
Willmer, C.~N.~A., da Costa, L.~N., \& Pellegrini, P.~S.\ 1998, \aj, 115, 
869 

\bibitem[Yan, Thompson \& Soifer (2004)]{yan04}
Yan, L., Thompson, D., \& Soifer, B.T., et al. 2004, \aj, in press

\bibitem[Zehavi et al.(2002)]{zeh02}
Zehavi, I., et al. 2002, \apj, 571, 172

\end{thebibliography}
\end{document}